\documentclass[sigconf]{acmart}

\usepackage[utf8]{inputenc}
\usepackage{subcaption}
\usepackage{caption}

\usepackage{enumitem}
\usepackage[textwidth=1cm,textsize=smaller]{todonotes}

\usepackage{soul}
\setstcolor{teal}
\usepackage{xcolor}

\newcommand{\lit}[1]{\textcolor{gray}{#1}}

\AtBeginDocument{%
  \providecommand\BibTeX{{%
    \normalfont B\kern-0.5em{\scshape i\kern-0.25em b}\kern-0.8em\TeX}}}

\copyrightyear{2021}
\acmYear{2021}
\setcopyright{acmcopyright}\acmConference[FAccT '21]{Conference on Fairness, Accountability, and Transparency}{March 3--10, 2021}{Virtual Event, Canada}
\acmBooktitle{Conference on Fairness, Accountability, and Transparency (FAccT '21), March 3--10, 2021, Virtual Event, Canada}
\acmPrice{15.00}
\acmDOI{10.1145/3442188.3445904}
\acmISBN{978-1-4503-8309-7/21/03}

\begin{document}

\title{An Agent-based Model to Evaluate Interventions on Online Dating Platforms to Decrease Racial Homogamy}
\title[\resizebox{4.2in}{!}{An Agent-based Model to Evaluate Interventions on Online Dating Platforms to Decrease Racial Homogamy}]{An Agent-based Model to Evaluate Interventions on Online Dating Platforms to Decrease Racial Homogamy}

\author{Stefania Ionescu}
\email{ionescu@ifi.uzh.ch}
\orcid{}
\affiliation{%
  \institution{University of Zürich}
  \streetaddress{Andreasstr. 15}
  \city{Zürich}
  \state{Switzerland}
  \postcode{8050}
}

\author{Anikó Hannák}
\email{hannak@ifi.uzh.ch}
\orcid{}
\affiliation{%
  \institution{University of Zürich}
  \streetaddress{Andreasstr. 15}
  \city{Zürich}
  \state{Switzerland}
  \postcode{8050}
}

\author{Kenneth Joseph}
\email{kjoseph@buffalo.edu}
\orcid{}
\affiliation{%
  \institution{University at Buffalo}
  \streetaddress{Davis Hall}
  \city{Buffalo}
  \state{NY, USA}
  \postcode{14260}
}

\renewcommand{\shortauthors}{Trovato and Tobin, et al.}

\begin{abstract}
 Perhaps the most controversial questions in the study of online platforms today surround the extent to which platforms can intervene to reduce the societal ills perpetrated on them. Up for debate is whether there exist \emph{any} effective and lasting interventions a platform can adopt to address, e.g., online bullying, or if other, more far-reaching change is necessary to address such problems. Empirical work is critical to addressing such questions. But it is also challenging, because it is time-consuming, expensive, and sometimes limited to the questions companies are willing to ask. To help focus and inform this empirical work, we here propose an agent-based modeling (ABM) approach. As an application, we analyze the impact of a set of interventions on a simulated online dating platform on the lack of long-term interracial relationships in an artificial society. In the real world, a lack of interracial relationships are a critical vehicle through which inequality is maintained.  Our work shows that many previously hypothesized interventions online dating platforms could take to increase the number of interracial relationships from their website have limited effects, and that the effectiveness of any intervention is subject to assumptions about sociocultural structure. Further, interventions that are effective in increasing diversity in long-term relationships are at odds with platforms' profit-oriented goals. At a general level, the present work shows the value of using an ABM approach to help understand the potential effects and side effects of different interventions that a platform could take.
\end{abstract}



\vspace{-5em}

\maketitle

\section{Introduction}

Online platforms have been criticized for their role in a profusion of societal ills, from the increased spread of political misinformation \cite{benham2012web}, to extrajudicial killings in the Philippines \cite{pimentel_facebook_nodate}.  Platforms have sought to address these issues with various \emph{interventions}, from attempts to filter or restrict various forms of content (e.g. hate speech) \cite{hern_how_2020} to attempts to decrease discriminatory practices by users \cite{dickey_heres_2016}. Few, if any, would argue that these interventions have fully addressed the problems at hand.  Underlying agreement to this high-level assertion, however, lies a more debated question--\emph{what are the limits of a platform's ability to intervene on society's ills?}

Given the documented problems platforms have helped give rise to, it would be wrong to assume that they have no role to play in positive change. At the same time, however, we must be careful to ascribe too heavy of a causal arrow to platforms.  Doing so is destined to lead to an ill-informed aura of technological determinism \cite{marx1994does}, where we search for solutions to societal problems by endlessly tweaking the structure of our technology, ignoring other root underlying causes \cite{fisman2016fixing}.

How might we understand the limits to a platform's ability to intervene to reduce a given societal ill? Empirical work, especially based on natural experiments \cite{malik2016identifying} and using algorithmic auditing \cite{sandvig2014auditing,hannak2013measuring}, is critical, but is restricted to studying the platform as it currently exists. Exploration of the effects of interventions is thus limited to what platforms are willing and able to implement given their profit-oriented goals. Even beyond these profit-related restrictions, there are ethical questions about the use of experimentation without the consent of platform users \cite{siroker2013b}.

Even still, scholars have pursued ethical empirical research not bound by profit-oriented goals, for example, by constructing platforms of their own \cite{chen2017gender,may2019gender}. But doing so is time-consuming and expensive; it would be useful if we could first scope potential solutions and their potential side effects in a more rapid and expansive way. Social theory can partially fulfill this role \cite{joseph_theory}, but it can be difficult to theoretically characterize, and much less operationalize, the myriad causal factors underlying sociotechnical systems.

What is needed is a methodology that exists in this middle ground between difficult empirical work on the one hand, and often abstract theory on the other.  Simulation has long filled this role of bridging theory and empirical work in the social sciences \cite{schelling_dynamic_1971,gilbert_simulation_2005,epstein_agent-based_1999}, and more recently has proved useful in the study of algorithmic fairness. For example, \cite{lum_predict_2016}, and later \cite{ensign_runaway_2017}, use simulation to show concretely how the abstract notion of feedback loops in predictive policing can lead to settings where police unfairly target one population over another, despite equal crime rates. Simulation has the added benefit of being applied to hypothetical people, allowing us to carefully and fully consider the harms that may be caused by a particular intervention before it has any real effects.In the present work, we use simulation, and specifically, an \emph{agent-based model (ABM)}, to make concrete a set of proposed interventions to address a given societal ill on a given online platform, and to provide new ideas about their implications and side-effects.

\paragraph{Online Dating} Specifically, we construct an ABM of an online dating platform and assess how interventions theorized recently by Hutson et al. in their award-winning CSCW paper \cite{hutson2018debiasing} affect the degree of racial homogamy in a simulated society. \emph{Racial homogamy} is defined as the \emph{tendency of individuals to seek out and engage in long-term relationships with individuals who are the same race}.

Critically, we do not argue that racial homogamy is in and of itself a societal ill; there are many valid reasons why individuals, particularly those identifying with an oppressed racial group, may want to maintain racial homogamy in their relationships. However, racial homogamy has been well-established to contribute to the reinforcement of racial inequality by increasing racial segregation \cite{baldassarri2007dynamics} and racialized income inequality \cite{grodsky2001structure}.  \citet{hutson2018debiasing} argue that racial homogamy is hence worthy of dating-platform intervention, as it is one racializer of economic and social inequality \cite{hutson2018debiasing}.

After detailing this connection between racial homogamy and racial inequality, \citet{hutson2018debiasing} explore an array mechanisms within online dating platforms that present opportunities for interventions to reduce racial homogamy. For example, ``search, sort, and filter tools'' allow users to perform initial searches for romantic partners, and thus might encourage users to make particular decisions about the race of their partner.  
Hutson et al. therefore take, at least implicitly, the stance that platforms could construct interventions that reduce racial homogamy.  Others, however, dispute this claim. In particular, \citet{anderson2014political} study racial bias in online dating from a sociological perspective, 
and conclude that ``merely altering the structural environment [on the platform]—say, by creating more opportunities for individuals of different races to interact—would not necessarily ameliorate persistent patterns of racial homogamy in romantic relationships'' (pp. 38-39). This claim puts the ideas presented in Anderson et al.'s work in contrast with those presented by Hutson et al., who explicitly argue that structural changes to platforms can reduce racial homogamy. 

These two papers therefore present a difference of opinion on the extent to which interventions on online dating platforms can impact racial homogamy, and ultimately racial inequality. However, empirical analyses of this debate is difficult for the reasons listed above. We therefore employ a simulation model to bridge the middle ground between \citet{hutson2018debiasing} theorized interventions and potential future empirical work. Our work helps to operationalize the theoretical constructs and to explore their impacts and potentially unintended consequences in an \emph{artificial}, but \emph{empirically and theoretically grounded}, complex sociotechnical system.

\paragraph{Contributions} The present work makes three contributions:
\begin{itemize}
	\item  We provide a blueprint (including an open-source modeling framework)\footnote{Repository link:  https://github.com/StefaniaI/ABM-for-Online-Dating.git.} for the use of agent-based modeling to help operationalize theory and explore the potential impacts of interventions of online platforms
	\item Specific to racial homogamy and online dating platforms, we find that various interventions theorized by \citet{hutson2018debiasing} indeed decrease racial homogamy in our simulated society according to our selected measure by as much as 10\%
	\item \emph{However}, the benefits of these interventions are dependent on specific assumptions about the sociocultural structure of that society, and are often at odds with societal outcomes and profit-oriented goals of the platform
\end{itemize}

Like many social science endeavors \cite{watts2011everything}, our core findings present claims that, once stated, seem quite obvious. However, given their motivation from established differences of opinion in published literature by leading scholars, our work serves a critical role in the path forward to understanding the potential of platforms to intervene to reduce racial homogamy, and as a broader methodology to address similar questions in other domains.

\section{Related Work}

\subsection{Online Dating}
Online dating plays an increasingly central role in the process of finding romantic partners, with 15\% of Americans reporting to have used an online dating site \cite{Cacioppo10135}. Indeed, at least for heterosexual couples, online dating has become the predominant means of meeting romantic partners \cite{rosenfeld2019disintermediating}. This change in the means for dating has opened the door to a debate on whether platforms can disrupt patterns of assortative mating, and in particular decrease racial homogamy. While \citet{hutson2018debiasing} argue that platforms have the means of decreasing racial homogamy through interventions, \citet{anderson2014political} think that this would likely not be the case as homogamy is partly caused by same-race preferences which cannot be changed through opportunity alone. These preferences influence and are influenced by social norms, and their formation is based on imitation and recurrent behaviour \cite{opp1982evolutionary} and enforced through social verification \cite{hardin1996shared}. To capture this phenomena, we consider norms and their impact on preferences. In alignment with \citet{hutson2018debiasing}, we acknowledge that platforms can potentially intervene to change norms on the platform, so, in our model, we differentiate between offline and online norms.

The process people go through while dating online is complex in multiple ways. First, one goes through a variety of different ``phases'' of dating, from seeking information about dating platforms to developing a long-term relationship \cite{finkel2012online}. Second, the decision-making process depends on distinct factors such as the perceived level of compatibility \cite{bruch2018aspirational}, how much we can validly infer about others online \cite{frost2008people}, and one's own personal stereotypes \cite{krueger1996personal}. 

The complexity of studying dating have naturally given rise to the use of simulations. Previous agent-based models for dating have looked at, for example, how the selection of partners is influenced by space and mobility \cite{smaldino2012human}, and whether assortative mating patterns could explain previously observed statistics on marriages and divorce \cite{hills2008population}. We build upon these examples in our work, but focus on the novel problem of understanding the role of platform interventions on racial homogamy in the context of online dating.




\subsection{Uses of Simulation to Study Fairness}

Starting with the seminal work of \citet{schelling_dynamic_1971} on housing segregation, simulation has been widely leveraged to explore how inequality and unfairness can arise and be combated in social systems. 
These studies have provided useful fodder for the more recent application of simulation to questions of \emph{algorithmic} fairness. For example, \citet{martin_jr_extending_2020} use a type of simulation called System Dynamics Modeling \cite{karnopp2012system} to show how simulation can help us to think through the complex impacts of different interventions in the context of credit lending. Others have leveraged different kinds of simulation models to study lending in similar ways, for example, within a reinforcement learning paradigm \cite{damour_fairness_2020}.
Still others have used simulation beyond the context of lending. For example, various forms of simulation have been used to study fairness and feedback loops in the context of predictive policing (as noted above \cite{lum_predict_2016,ensign_runaway_2017}) the effects of recommender systems \cite{bountouridis_siren_2019}, and income inequality for drivers on rideshare platforms \cite{bokanyi2020understanding}. 

Like \citet{bokanyi2020understanding}, we use an agent-based model here.  Agent-based modeling involves the construction of virtual worlds in which artificial agents interact according to a set of pre-specified rules \cite{epstein_agent-based_1999}.  While behavior is thus predictable at the individual level, ABMs are useful for understanding how these pre-specified micro-level behaviors result in emergent properties, e.g., social biases across social groups \cite{joseph_coevolution_2014} or opinion polarization \cite{schweighofer2019weighted}.

Our work compliments existing literature using simulation modeling to study questions of fairness and equality in three ways. First, we study a concrete, contemporary, and contested question about the role that online dating platforms can play in reducing racial homogamy.  Second, rather than focus on algorithmic fairness specifically, we emphasize other ways in which online platforms can have impacts on society.  Finally, and at the same time, we develop an agent-based model that can be easily extended to study algorithmic fairness, complementing existing System Dynamics and reinforcement learning platforms provided by prior work \cite{damour_fairness_2020,martin_jr_extending_2020}.

\section{Model}

Our model contains a variety of parameters, listed in Table~\ref{tab:virt_exp} alongside references, where applicable, to were used to determine their values. Below, we first provide an overview of the model, and then further details on each major element of the model.

\subsection{Model Overview}

\begin{figure}[t]
\centering
\begin{subfigure}{.45\linewidth}
  \centering
  \includegraphics[width=\linewidth]{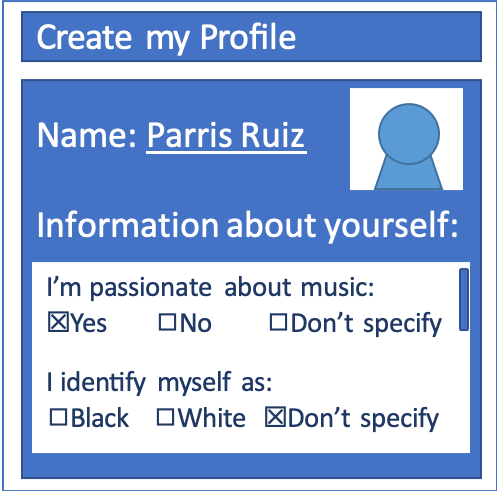}  
  \caption{Creating a profile}
  \label{fig:create_prof}
\end{subfigure}
\hfill
\begin{subfigure}{.45\linewidth}
  \centering
  \includegraphics[width=\linewidth]{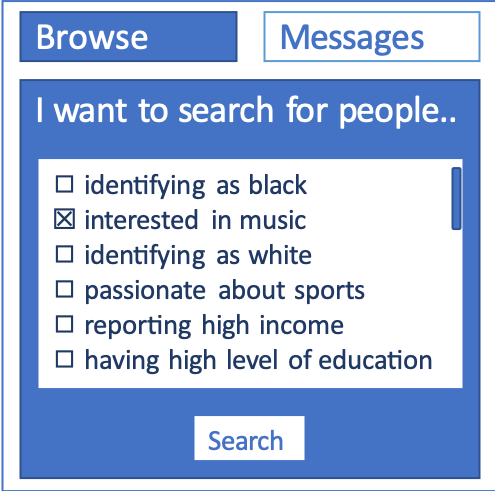}  
  \caption{Searching (using filters)}
  \label{fig:filtering}
\end{subfigure}

\begin{subfigure}{.45\linewidth}
  \centering
  \includegraphics[width=\linewidth]{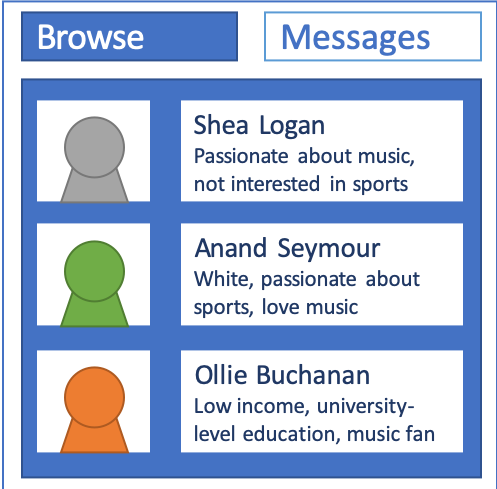}  
  \caption{Browsing the (filtered) list of potential partners}
  \label{fig:browsing}
\end{subfigure}
\hfill
\begin{subfigure}{.45\linewidth}
  \centering
  \includegraphics[width=\linewidth]{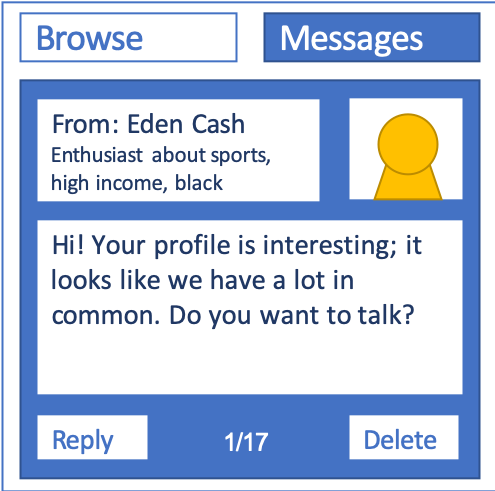}  
  \caption{Looking at (and potentially responding to) messages}
  \label{fig:sub-second}
\end{subfigure}

  \caption{The actions agents can take on the platform}
\label{fig:dating_platform}
\end{figure}

\begin{figure*}[t]
\centering
	\includegraphics[width=\linewidth]{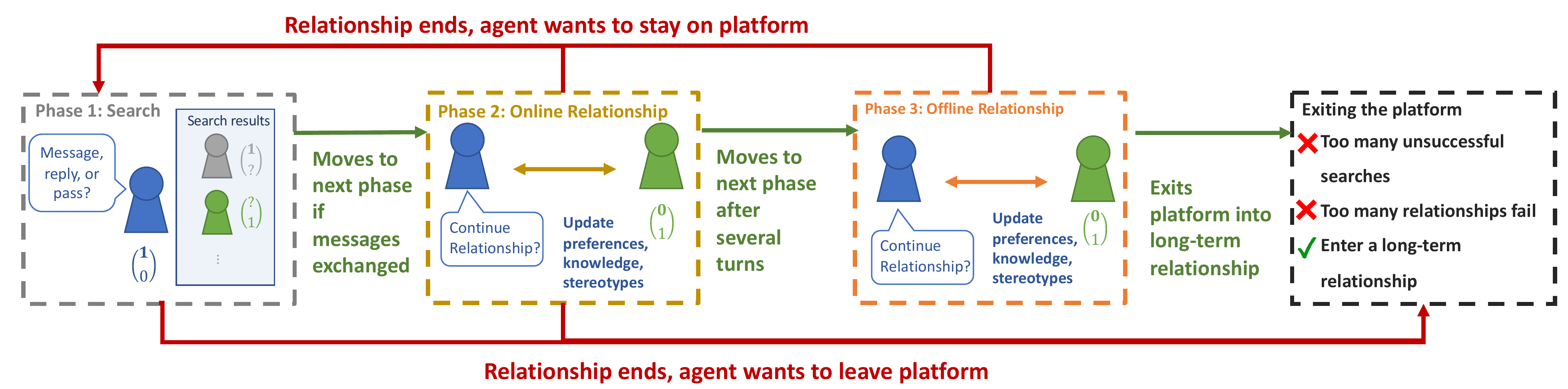}
	\caption{Overview of relationship phases and transitions in each turn of an iteration, explained through the blue agent's choices and actions. Quote boxes represent decisions the blue agent makes, and the 0/1 vectors represent the agent's attributes, or the blue agent's knowledge of potential partner's attributes.}
	\label{fig:rel_phases}
\end{figure*}

\begin{table}[ht]
\caption{\label{tab:virt_exp} Tabular description of model parameters (Left) and the values taken in the virtual experiment (Right). For parameters with a red-coulured value, we simulate each value for all simulated interventions but keep the values of the other parameters to red values  (see Section~\ref{sec:virtual_experiment}).}
\small
\begin{tabular}{p{5.7 cm} p {2.25 cm}}
\hline
Parameters & Values Taken \\
\hline
{\bf Varied in Virtual Experiment}  & \\
\ \ \  Tolerance to un-successful search turns & 20/ {\color{red} 25}/ 30\\
\ \ \  Out-platform norms for protected attributes & 0/0.1/ {\color{red} 0.2}/0.3\\
\ \ \ Strength of updates: norms $\to$ pref.s  &  0/ {\color{red} 1\%}/ 5\%\\
\ \ \ Initial degree of ethnocentrism & 0/ {\color{red} 0.25}/ 0.5/ 1\\
\ \ \ Correlations in attribute generation: $\beta, \gamma$ & 0/ {\color{red}0.4}/0.6/0.8, {\color{red} 0.2}/0.6\\
\hline
{\bf Fixed in Virtual Experiment} & \\
\ \ \ Strength of updates: interaction $\to$ pref. &  2\%\\
\ \ \ Strength of updates: average pref. $\to$ norms  &  5\%\\
\ \ \ Relative importance of out-platform norms when generating initial preferences &  20\%\\
\ \ \ Failure tolerance to unsuccessful relationships  & 20\\
\ \ \ \# initial acquaintances in-group \& out-group & 300, 100 \lit{\cite{mcpherson_birds_2001,tajfel_social_1971,joseph_coevolution_2014}} \\
\ \ \ Weight of searchable vs experiential  attributes & 1 - 3 \lit{\cite{frost2008people}}\\
\ \ \  Pr. searchable attribute specified on profile & 50\% \lit{\cite{lewis2016preferences}}\\
\ \ \  Pr. learning searchable \& experiental attributes & 50\% \lit{\cite{finkel2012online}}\\
\ \ \  \# rounds until offline \& long-term & 7, 37 \lit{\cite{ramirez2015online, brym2001love}}\\
\ \ \  Pr. to meet while offline & 14\%\lit{\cite{brym2001love, munoz2007aggression}}\\
\ \ \  Pr. to consider a message (while in search) & 50\%\lit{\cite{finkel2012online}}\\
\ \ \  Maximum \# agents considered in a search turn & 30 \lit{\cite{frost2008people, finkel2012online}}\\
\ \ \  \# iterations & 2000\\
\ \ \  Initial population size & 300\\
\ \ \  \# agents added per iteration & 4 \\
\hline
\end{tabular}
\end{table}

\emph{Agents} in our model are people that have at some point used our simulated dating website. Agents are assumed to carry out a series of activities on the site in pursuit of a romantic partner. These activities include creating a profile (Figure~\ref{fig:dating_platform}a), searching for a partner (b), browsing the list of potential partners returned by that search (c), and messaging (d). Agents in our model perceive other agents according to their \emph{attributes}.  As in most ABMs \cite{gilbert_agent-based_2007}, attributes are just a vector of 0s and 1s, but are assumed to represent real-world traits that might be important in the context of dating, e.g. partisanship. All agents also have a ``protected attribute,'' that is, an attribute relevant to questions about discriminatory dating practices. We will focus on race as the protected attribute here. 
Moreover, we will focus on a binarized model of race, that is, agents can be only one of two races in the model. In the context of prior work we discuss above, the two races of interest here are White and Black. Section~\ref{sec:agents} provides full details on how agents are modeled.
 
As shown in Figure~\ref{fig:rel_phases}, relationships between agents evolve over a series of \emph{phases}, from finding (or being found by) a potential partner, to interacting with each other on the platform via messages (online relationship), and finally to an offline relationship \cite{finkel2012online}. At any phase, a relationship might end, at which point the agents will resume their search for another romantic partner, or exit the platform. If a relationship continues for long enough, it becomes a \emph{long-term} relationship. Our primary outcome measures focus on racial homogamy in these long-term relationships. and Section~\ref{sec:phases} provides complete details on the different relationships phases. 

The simulation is carried out via a sequence of iterations, or \emph{turns}, with one turn being loosely the equivalent of one day. On the first turn of the simulation, 300 agents are initialized and become the population of users on the online dating site. On each subsequent turn, we add four new agents, modeling a constant flow of new users.  After new agents are added, all agents get a turn to act. During their turn, agents can 1) conduct a search, potentially using a set of search \emph{filters} provided by the site, 2) send a message, 3) start a new relationship, 4) continue or end an existing relationship, or 5) leave the platform, either because they have found a long-term partner, or because they have become frustrated with the site.  

As shown in Figure~\ref{fig:rel_phases} , which of these things agents decide to do is based on which part on the relationship phase they are in, and on the \emph{decision model} we construct for agents. The decisions made by an agent are determined by 1) their attributes, 2) the attributes of their potential (or current) partner, 3) the \emph{stereotypes} they hold of how different attributes are related, and 4) their \emph{preferences} for others with certain attributes. These preferences are themselves informed by \emph{social norms} that exist both on and off the platform. Section~\ref{sec:decisions} provides details on how agents make decisions. 

Agent's attributes are static in our model. However, their stereotypes and preferences \emph{evolve} as they interact with others and move through relationship phases.  The social norms governing all agents also evolve throughout the simulation. Section~\ref{sec:bel_updates} details when and how these updates to stereotypes, preferences, and norms occur.  

\subsection{Agents}
\label{sec:agents}
As in all ABMs, our agents have state and can take actions. Agents' states encompass their \emph{type}, \emph{attributes}, \emph{stereotypes}, \emph{preferences}, and \emph{knowledge of others}. The actions agents take, based on these states, are a set of \emph{decisions}, which we describe in Section~\ref{sec:decisions}.

\subsubsection{Type} 
Agents can be of two types, and are only attracted to agents of the opposite type. We thus loosely mimic a hetero- and mono-normative online dating website, as was studied by \citet{anderson2014political}. We return to this limitation in our conclusion. Agents are assigned to the two modeled types with equal probability.

\subsubsection{Attributes}

Each agent $i$ has a vector of binary attributes, $a^{(i)}$. We distinguish between race, $a^{(i)}_0$, and $K$ other attributes. Following theory in the online dating literature \cite{finkel2012online,bruch2018aspirational}, we make two additional distinctions among attributes. First, we distinguish between \emph{matching} and \emph{competing} attributes. For \textit{matching} attributes, agents try to find partners with similar values, while for the \textit{competing} attributes, they try to find partners with the highest possible values. We assume that the protected attribute is matching. 

Second, we distinguish between attributes visible only offline as \emph{experiential} attributes, and ones that are visible both online and offline as \emph{searchable} attributes. Examples of searchable attributes include personal interests and location. Sense of humor is an example of experiential attribute \cite{finkel2012online}. Since people do not share everything about themselves in their profiles \cite{lewis2016preferences}, agents have a probability of $50\%$ to specify each searchable attribute on the platform. In the example from Figure~\ref{fig:dating_platform}, the agent chose to specify their music taste, but not their race.
We describe the complete process of generating attributes for new agents in the Supplementary Materials. 


\subsubsection{Stereotypes} 

Agents' \emph{stereotypes} encompass their beliefs of \emph{which attribute combinations are more likely to be encountered} \cite{krueger1996personal}. We model the stereotypes of agent $i$ as a matrix $G^{(i)}$.  This matrix is initialized via a theoretically grounded model that is also based on prior work. Specifically, when generating a new agent, we first generate a sample of $300$ agents having the same race, and $100$ agents with a different race. The imbalance in sample sizes is motivated by homophily: people are, in general, more likely to meet and know people similar to themselves \cite{mcpherson_birds_2001}. In addition to homophily, we also model in-group favoritism, i.e. that people tend to look favorably on people like them \cite{hastorf_they_1954,tajfel_social_1971}. To capture this in our model, we follow prior agent-based models \cite{joseph_coevolution_2014} and assume that agents also have an initial degree of ethnocentrism. For example, as in \cite{joseph_coevolution_2014}, if the degree of ethnocentrism is $0.3$, then the agent will, on average, expect 30\% of the attributes of agents in the other-race sample to have values that the agent does not like.

The Supplementary Materials contain full details on stereotype initialization, and Section~\ref{sec:bel_updates} on how stereotypes evolve over time.

\subsubsection{Preferences} 

An agent's \emph{preferences} define which attributes they believe are most important in a potential partner. We model the preferences of agent $i$ as a weight vector, $w^{(i)}\in \Delta^{K}$. This vector of preferences is in the $K$th simplex, i.e. its entries sum to one, thus capturing the \emph{relative} importance of different attributes in the eyes of the agent. For example, a weight vector with $w^{(i)}_j = 1$ and all other entries of $0$ corresponds to an agent who only considers the $j$th attribute to be important in a potential partner.

The preferences of agents influence and are influenced by two different types of social norms, i.e. sociocultural regulations on one's preferences \cite{lizardo2010skills}.\footnote{We discuss initialization of preferences and norms in the supplementary material.} First, while interacting online, an agent impacts and is impacted by \emph{platform norms}. 
Second, when in an offline relationship, the agent preferences inform and are informed by off-platform norms. Critically, then, in our model as in the real world \cite{lizardo2010skills}, agent's preferences thus vary in part based on social norms, which change as they move between the context of the platform and their offline social life.  At the same time, norms are aggregations of the preferences of individuals, except, as we will discuss, when explicit efforts are made by the platform to fix norms.

Agent preferences are initialized based on a procedure common in the decision-making literature \cite{cantwell2020thresholding}. We note here that two parameters govern important aspects of this process. First, a parameter $\beta$ is used to capture the level of correlation between race and other attributes. Second, $\gamma$ captures the level of correlation between the non-protected attributes. Varying these two parameters changes the predictability of one attribute, given another. In our virtual experiment, both $\beta$ and $\gamma$ are changed. Most importantly, $\beta$ is varied to reflect different degrees to which race can be inferred by agents based on other available attributes when the agent chooses not to display race in their online profile.


\subsubsection{Knowledge of other agents} 

As they interact with another agent, agents can become aware of the values of new attributes of that agent.  This knowledge is used to make decisions about, e.g., whether to continue relationships. How agents acquire this knowledge is discussed in Section~\ref{sec:phases}.

\subsection{Relationship Phases}
\label{sec:phases}
Agents can be in one of three relationship phases: 1) searching, 2) in an online relationship, or 3) in an offline relationship.\footnote{We make the simplifying assumption that an agent can only be in one of the before-mentioned phases at one time, e.g. one cannot be both searching and dating offline.}

\subsubsection{Searching phase}
When first entering the platform, an agent is in the searching phase. An agent can search in two ways. First, the agent can perform a query for potential partners using a search interface. Second, the agent can look through messages in their inbox that they have received from other agents.  
In one turn of the simulation, agents can perform up to $30$ search actions, mimicking the typical length of a user's session on an online dating website \cite{frost2008people, finkel2012online}. A search action represents looking at either a single search result or a single message in the agent's inbox. The agent randomly decides which of these two actions to take. If the agent decides to view a profile from their search result list, they look at the profile of the next search result and decide whether or not they should send that person a message. If the agent decides to view a message in their inbox, they look at the oldest un-read message and consider whether or not they should reply. 

Search results are a randomly ordered subset of other agents currently on the platform. Importantly, however, the agent \emph{decides which other agents will appear in their search results}. They do so by \emph{filtering} for potential dating partners.  The agent filters on a single attribute, the one that, on this turn, is most important to them (i.e. for agent $i$, the attribute with the highest value of $w^{(i)}$). In Figure~\ref{fig:dating_platform}, for example, interest in music is the most important factor for the blue agent, and they hence use this as a filter. The blue agent is then only shown profiles of other agents having a passion for music. 

If the agent decides to reply to a message in their inbox, their search ends, and the two agents involved proceed to an online relationship phase. However, the agent can send as many initial messages to agents in their search results as they would like. An \emph{unsuccessful search} is considered to be a turn in which the agent is not interested in answering any of the messages they have looked at, and in which they did not find at least half of the profiles they've looked at interesting enough to send an initial message. Too many unsuccessful search turns would make the agent exit the platform.

\subsubsection{Online relationship phase}
After answering or getting a reply to an initial message, the agent moves to an Online Relationship Phase with another agent. While in this phase, in each iteration the two agents within the relationship have an online interaction. During this interaction, three things happen. First, agents may update their knowledge about the attributes of the other. Going back to the example in Figures~\ref{fig:dating_platform} and~\ref{fig:rel_phases}, when entering the relationship, the blue agent only knows the value of the green agent's interest in music. After one interaction, they have also learned the self-identified race of the green agent.  Second, agents update their stereotypes based on what they learn about the other; see Section~\ref{sec:bel_updates}. 
Finally, agents decide whether or not to continue the relationship.  If both decide to continue, the interaction is \emph{positive}, and the relationship will continue with a new interaction in the next iteration. Otherwise, the interaction is \emph{negative}, and the relationship ends.


After deciding on whether or not to continue the relationship, the preferences of agents are updated based on whether the interaction was positive or negative (see Section~\ref{sec:bel_updates}) and the turn of the agents ends. If the agents have decided to end the relationship, their \emph{tolerance to failed relationships decreases}. Agents will exit the platform if this tolerance reaches $0$, and otherwise they return to searching. If the agents decide to continue the relationship, they either continue the online relationship, or, if the relationship has lasted long enough, move to the Offline Relationship Phase.

\subsubsection{Offline relationship phase}
The turn of an agent in an Offline Relationship Phase progresses similarly to the Online one, with two key differences. 
First, if the interaction was positive, the agents either continue the relationship (as above); however, if the relationship lasts for $30$ iterations, the agents then exit the platform into a \emph{long-term} relationship. Second, while offline, in each iteration there is a fixed $1/7$ chance of interacting. This means that, in expectation, people in offline relationships meet once per week \cite{brym2001love, munoz2007aggression}.

\subsection{Agent Decision-Making}
\label{sec:decisions}

In the Searching phase, agents answer the questions (posed to themselves), "should I reply to/send an initial message to this agent?". When in a relationship, the agent answers questions of the type "should I continue being in a relationship with this agent?". Both of these are yes/no questions, and agents use the same decision function, described below, to answer them.

Suppose that agent $i$ must make a decision about sending a message or continuing a relationship with agent $j$. We assume agent $i$ may know only an incomplete vector of agent $j$'s attributes, which we call $\hat{a}^{(j)}$. We use $\hat{a}^{(j)} \subseteq a$ if $a$ extends the incomplete vector of attributes $\hat{a}^{(j)}$, i.e. $a$ is a complete vector that has the same entries as $\hat{a}^{(j)}$ on the known attributes. Lastly, for the agents $i, j$ there is some time value, $t$, capturing the amount of time spent in that specific phase with one another (for somebody in the searching phase or who just started an online or an offline relationship, $t=0$). 

In making decisions, agents consider a) the time spent so far with the other agent and b) how well the attributes of the other agent align with their own attributes and preferences. Time is an important predictor for breakups, as agents that have stayed longer in a relationship are less likely to interrupt their relationship. For example, people tend to be especially selective in their first offline date together \cite{long2010scripts}. Consequently, using the above introduced notation, the probability that agent $i$ answers yes to a decision question regarding agent $j$ is given by the following sigmoid function:$$\mathbb{P}(\text{yes}) = \frac{1}{1 + \exp \left(- \mathbb{E} \left[w^{(i)}\cdot s^{(i)}(a) |  \hat{a}^{(j)} \subseteq a \right]- t \right)}.$$

Two aspects of the above equation require further explanation. First, the expected value is taken over all possible attribute combinations extending the incomplete information of $i$ over the attributes of $j$. The probability of each combination is given by the current stereotypes of $i$. Thus, $i$'s stereotypes impact what they \emph{expect} the unknown attributes of $j$ to be. Second, the function $s^{(i)}(a)$ is an attribute scoring function that for a given agent and complete attribute vector returns a vector with entries $1$, $0$, and $-1$ depending on whether each attribute is viewed as good, neutral, or bad. Agents see a matching attribute value as being \textit{good} if it is the same as their own and \textit{bad} if it is different. Similarly, they see a competing attribute value as \textit{good} if it is larger than their own, \textit{neutral} if it is equal, and \textit{bad} if it is lower. The Supplementary Material provides a formalization of this scoring function.

\subsection{Belief, Preferences, and Norm Updates}
\label{sec:bel_updates}
Agents' stereotypes, preferences, and the norms of the platform change on each iteration. 

\subsubsection{Agent Stereotype Updates}

Recall that when generating an agent, we construct $G^{(i)}$, a sample of attribute combinations they have encountered so far. After each interaction, an agent observes a new combination of attributes. If that observation is a complete list of attributes, they add it to the sample. Otherwise, they find the probabilities of each possible complete combination of attributes that could extend the observed one and adds a proportional fraction to each of these categories. For example, consider somebody that observes the combination $(1, 0, \bot)$ after an interaction, and their sample was $25$ observations for $(1, 0, 0)$ and $75$ for $(1, 0, 1)$. Post-interaction, their sample changes to $25.25$ and $75.75$. 

\subsubsection{Norm and Preference Updates}
Preferences and on-platform norms both change during each iteration. First, agent preferences change according to their interactions based on a) whether the interaction was positive or negative and b) the attributes of agents. For example, a positive interaction with somebody of the same race will change the preference of the agent by increasing the preference of an agent to date someone of the same race. In contrast, a negative interaction with someone of the same race will make the agent \emph{less} likely to seek out a same-race partner in the future.

At the end of each simulation turn, norms and preferences are updated to reflect their convergence \cite{lizardo2010skills}. Specifically, the on-platform norm gets closer to the average preference of agents on the platform, and the preferences of agents are updated by getting closer to the respective norm. Each update is carried out by taking convex combinations between the original value of the norms (preferences) and the values of the influencing preferences (norms). The extent to which norms determine preferences, and vice versa, is set via model parameters. See the supplementary information for more details on how these updates are made, as well as details on initialization. 


\section{Virtual Experiment}\label{sec:virtual_experiment}

We now describe how we use our model to explore three different classes of \emph{interventions} outlined by \citet{hutson2018debiasing}.  In Section~\ref{sec:interventions}, we operationalize these interventions, which are also listed in Table~\ref{table:filter_inter}. We then discuss how we vary model parameters to evaluate the effectiveness of these interventions under different assumptions about the underlying society. Finally, we detail the outcome measures we use to understand the impacts of these interventions.  

\subsection{Interventions}
\label{sec:interventions}

\citet{hutson2018debiasing} suggest three places where platforms might intervene to reduce bias and discrimination on intimate platforms: 1) at the search, sort, and filter level, 2) at the matching algorithm level, and 3) at the community policy and messaging level. Here, we focus on the first and the last, and leave out the second for two reasons. First, not all platforms use matching algorithms, and even when they do, the algorithm differs substantially. Second, the literature on the possible algorithm to implement is vast~\cite{gale1962college, shapley1974cores, abdulkadirouglu2005boston, hitsch2010matching, tu2014online, xia2015reciprocal, brozovsky2007recommender}. 
 Consequently, there are many different alternatives to substitute the existing matching algorithm with. A comparison between these various  procedures is beyond the scope of this paper.


\subsubsection{Filtering Interventions}
The ease of filtering out people based on their attributes when searching for potential partners in online systems may exacerbate racial homogamy in two ways: 1) it reduces the diversity in search results and 2) it makes race a legitimate basis for decisions \cite{lewis2016preferences, hutson2018debiasing, bowker2000sorting}. The first suggested solution to mitigate this problem by \citet{hutson2018debiasing} is to purposefully introduce diversity into search results, e.g. by removing the ability to filter by certain attributes entirely. We refer to any intervention along these lines as a \emph{filter intervention}. We consider four different filter interventions, comparing them to a baseline of traditional platform behaviors.

As explained above, without any intervention, agents will filter their searches for romantic partners by the attribute they consider to be most important. We refer to this baseline option as \emph{Strong filtering}. From this baseline, \citet{hutson2018debiasing} note that interventions to reduce racial homogamy can be constructed along two dimensions. First, we can restrict the possibility of filtering on race entirely. To signal such an intervention we add a "non-race" label to the filter name. Second, we can introduce diversity by showing profiles of agents who did not provide the given attribute in their profile. We label filtering strategies that only include the desired setting of the given attribute as "Strong," and those that also include potential partners who simply do not fill out a value for this attribute as "Weak". Combining interventions on these two dimensions gives three more filtering options, namely Strong non-race, Weak, and Weak non-race. We also add a fourth intervention where no filtering is allowed at all, on any attribute, i.e. the "Off" filtering option.

The five filtering conditions are most clearly stated with an example. Consider the agents in Figure ~\ref{fig:dating_platform}. Let us assume the platform opts to use \emph{Strong} filtering. Further, let us assume that the blue agent's strongest preference is for a romantic partner that is the same race. In this case, the blue agent will search for others with the same race, and will see only other potential partners who publicly self-identified with that race on the platform.  If the platform had \emph{Weak} filtering, the agent would also see any other agents who a) have the same self-identified race or b) do not specify any race on the platform. If the platform employed \emph{Strong non-race} filtering, the platform would not allow the agent to search by race at all. The agent would therefore use their second strongest preference (say, music), to perform their search. Weak non-race follows similarly, and in the Off condition, agents would simply not be able to filter at all. Finally, note that we have described the process with a matching attribute; for competing attributes, agents are always assumed to search for other agents who hold the desired attribute.

\begin{table}[t]
\caption{The three types of Interventions, the baselines for each, and the different intervention conditions we test.}
\label{table:filter_inter}
\small
\begin{tabular}{p{1.8cm}  p{6.2cm}} 

{\bf Intervention} & {\bf Description} \\ 
 \hline\hline
 \multicolumn{2}{c}{{\bf Filtering Interventions}} \\
\emph{Strong} & \emph{Baseline. Only see others with the preferred attribute value on the most important attribute.} \\ 
Strong non-race & Strong filtering, if not on race. Otherwise, Strong filter on second most important.\\
Weak & Also shows profiles with unknown values on most important attribute\\
Weak non-race & Weak filtering, but not on Race.\\
Off & No filtering allowed. \\ \hline
 \multicolumn{2}{c}{{\bf Attribute Intervention}} \\
\emph{5 attributes} & \emph{Each agent has the protected attribute, plus one of each kind of attribute} \\
9 attributes& Each agent has the protected attribute, plus {\bf two} of each kind of attribute \\ \hline
 \multicolumn{2}{c}{{\bf Norm Intervention}} \\
\emph{Off} & \emph{Baseline. On-platform norms are an average of agent preferences} \\
On & On-platform norms are fixed to 0 on race, encouraging agents not to consider it\\
\end{tabular}
\end{table}

\subsubsection{Attribute Intervention}
Removing filtering options can make it harder for agents to find the partners they desire. A suggested solution by \citet{hutson2018debiasing} to mitigate this problem is to add or give additional attention to new non-race attributes on which agents can filter (e.g. political view or smoking preferences). We refer to this intervention as an \emph{Attribute Intervention}. In our model, this intervention is implemented by increasing the number of dating-relevant attributes from a baseline of 5 to a total of 9, i.e. from having one non-racial attribute of each type, to having two of them.\footnote{As a reminder, attributes differ along two dimensions matching-competing and searchable-experiential, thus giving 4 possible types. Note also that \citet{hutson2018debiasing} additionally suggest a variant of this intervention, namely one based on the minimal group paradigm in social psychology \cite{tajfel_social_1971}. We provide results for this similar intervention in the supplemental material.}


\subsubsection{Norm Intervention}

The final intervention from \citet{hutson2018debiasing} that we model are attempts to change or outline platform norms. \citet{hutson2018debiasing} suggest that different community policies, such as introducing pledges, educating users on the effect of expressing racial preferences, and detailing the disparities faced by racial minorities, can encourage social norms to date interracially. Norms, in turn, are known to be perhaps the most powerful mechanism for changing discriminatory behaviors \cite{tankard2016norm}. In our experiment, the platform norm intervention fixes the value of the on-platform norm for the race attribute to $0$. This means  that agents, via the link between norms and agent preferences, are encouraged not to consider race at all in their decisions. Differently from the baseline condition, when on-platform norms are updated to reflect the average of preferences, the value on the race dimension does not change and continues to be 0 (see the Supplemental Material). 
\subsection{Assumptions about Society}
\label{sec:society}

Table~\ref{tab:virt_exp} lists the model parameters we vary or keep constant in our virtual experiment.
As we will show, consideration of these parameters is critical to understanding how well claims about intervention effects generalize to different assumptions about society. At the same time, a full exploration of the parameter space is infeasible. We take a three-pronged approach to addressing this. First, as noted in the table, many of these parameter and their values are theoretically grounded and/or empirically informed by prior work. Second, we engaged in sensitivity testing, described in the appendix, across a range of values for all parameters. Finally, for our virtual experiment, we fixed parameters that did not appear to have significant impacts on model outcomes. For parameters that did appear to have an effect, we select a default value for each (coloured in red in Table~\ref{tab:virt_exp}); when varying one such parameter, we keep the others at their default values. 

As shown in Table~\ref{tab:virt_exp}, we vary five model parameters in our simulation. First, tolerance to unsuccessful search turns, which allows us to see how results change if agents are more or less tolerant to seeing undesired profiles and messages while browsing. Second, we vary racial bias in the out-platform norms, capturing how much importance society puts on same-race relationships compared to the other attributes of individuals. Third, we vary the influence norms have on preferences, i.e. the percentage of individual preferences explained by social norms. Fourth, we alter the initial degree of ethnocentrism, which intuitively corresponds to the extent to which agents are likely to prefer same-race relationships. The last parameters we change are the levels of correlation between attributes, i.e. $\beta$ and $\gamma$; we test in turn scenarios with zero, low, mild, and high correlations between race and the other attributes ($\beta$), and low and medium correlations between the other non-race attributes ($\gamma$).
Each simulation runs over $2000$ iterations, which corresponds roughly to a period of over $5$ years. Moreover, we use $20$ distinct random seeds for each scenario, thus having a total of $3000$ runs.

\subsection{Outcomes}
\label{sec:outcomes}

We treat racial \emph{hetero}gamy--the opposite of racial homogamy---as the main outcome variable when studying the impact of the simulated interventions on racial homogamy.\footnote{We do so since we would like to emphasize the positive nature of the proposed change.} We operationalize racial heterogamy as the percentage of long-term relationships that are interracial. However, a proportion-based measurement can of course vary based on changes in the numerator or denominator. Thus, an increase in racial heterogamy might be caused by a decrease in efficiency, i.e. the overall number of long-term relationships, or an increase in the number of interracial relationships. We therefore investigate these two quantities separately as well.

Lastly, we provide two measures of user satisfaction.
, which are important for two reasons. First, profit-oriented decisions by platforms depend in part on the number and, hence, also on the happiness of users. This in turn suggests platforms may be less likely to implement interventions which threaten these metrics. Second, even if a platform was somehow persuaded to implement such interventions, the free nature of the market permits unsatisfied users to easily move to a different dating platform.
To capture user satisfaction, we use two measures informed by literature on online dating \cite{frost2008people}. First, we measure the percentage of time spent in an offline relationship, relative to online dating phases. Second, we measure the number of users who exit the platform because they have exhausted their tolerance for unsuccessful online searches.

\section{Results}

\subsection{Average Effects on Racial Heterogamy}

\begin{figure}[t]
    \centering
    \includegraphics[width=\linewidth]{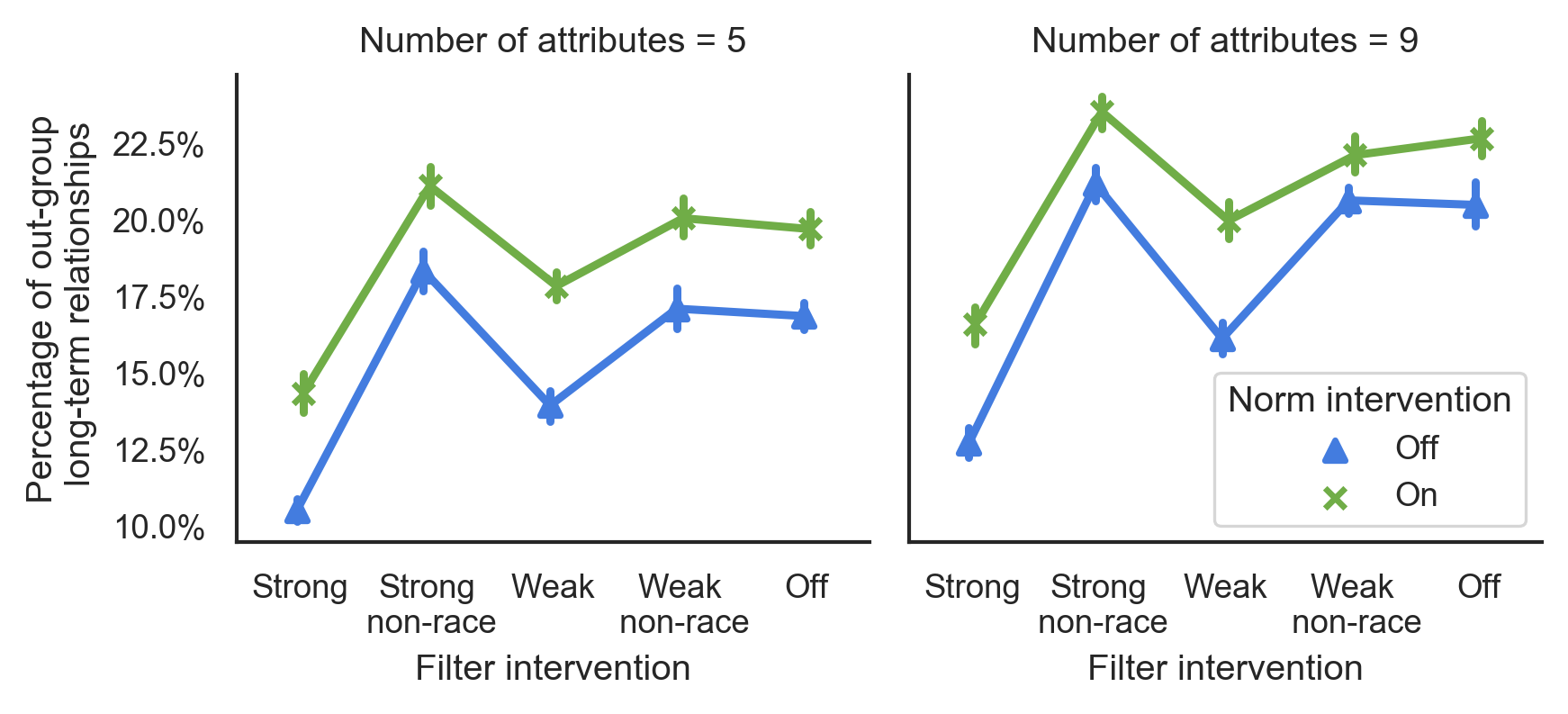}
    \caption{Racial heterogamy (y-axis) given a particular filter intervention (x-axis), and depending on whether or not the norm intervention is On (green lines, Xs as points) or Off (blue lines, triangles as points) and whether or not the attribute intervention is off (left subplot) or on (right subplot) }
    \label{fig:res_pr}
\end{figure}

Each intervention proposed by \cite{hutson2018debiasing} increases racial heterogamy. However, as shown in Figure~\ref{fig:res_pr}, the magnitude of these effects vary. The largest positive effect of any intervention is the Strong non-race filtering condition, i.e. when our simulated platform removes the option of filtering searches by race but allows agents to search on all other attributes. Doing so increases racial heterogamy from 10.5\% to 18.3\%. In contrast, a Weak filtering option increases the percentage by 3.4\% over the baseline.

Combining the Strong non-race filtering intervention with the norm and attribute interventions further increase racial heterogamy. However, the effects are non-additive. For example, the effect of the norm intervention decreases after implementing a filtering intervention. Applying the norm intervention to the baseline filtering option results in a 4\% increase in interracial relationships, while if it is applied on top of a Strong non-race intervention, the impact reduces to less than 3\%.  Finally, adding dating-relevant attributes produces an additional and relatively stable increase of 2\% to 3\%. These results suggest the potential of interventions on the platform, but also that their impacts may quickly reach a saturation point.

\subsection{Effects on Total Number of Relationships}

\begin{figure}[t]
    \centering
    \includegraphics[width=\linewidth]{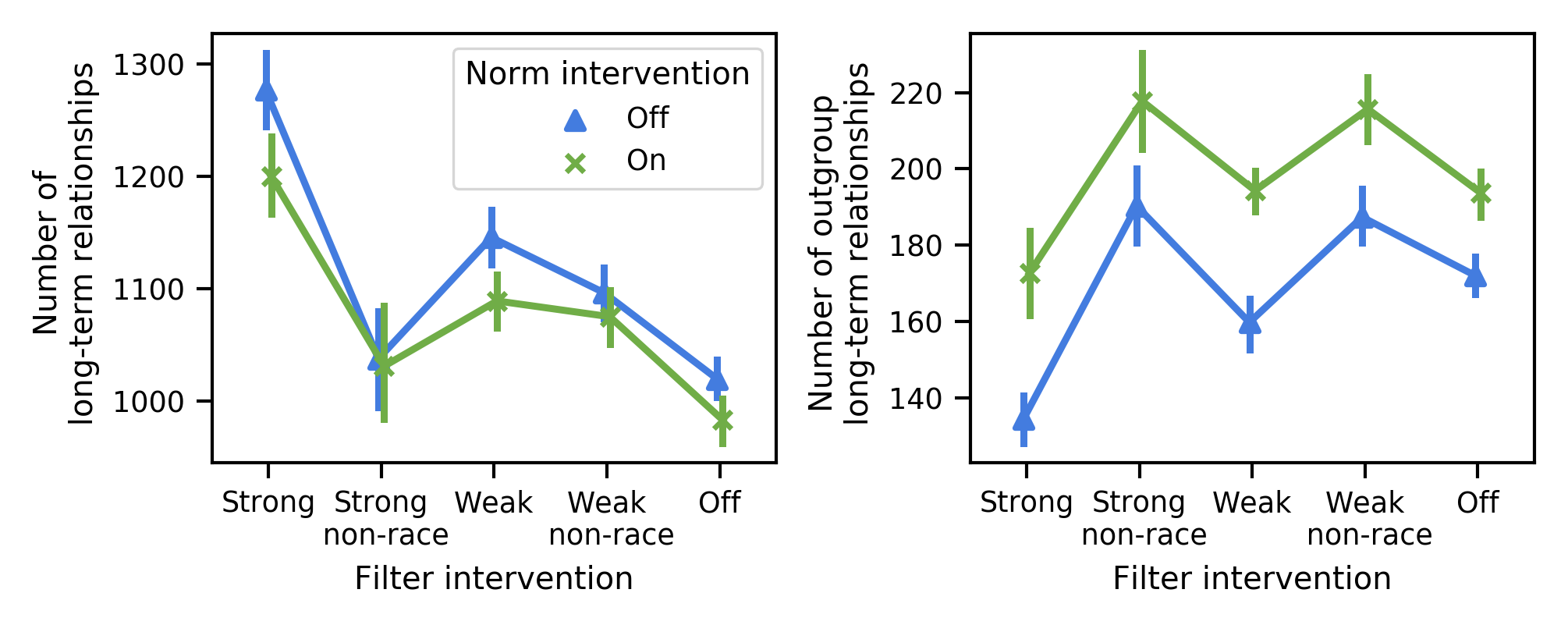}
    \caption{Number of total (left) and interracial (right) long-term relationships (y-axis) for the filter conditions (x-axis) and the norm intervention (Blue: off, Green: on). Results are only shown without the attribute intervention}
    \label{fig:res_no}
\end{figure}

We also find that increases in racial heterogamy--the \emph{percent} of long term interracial relationships--are not solely due to an increase in the \emph{number} of interracial long-term relationships. Rather, as depicted in the first part of Figure~\ref{fig:res_no}, most interventions are also paired with a significant decrease in the overall number of long-term relationships.

With respect to the filtering interventions, removing the ability to filter by race causes an almost 20\% drop in the total number of long-term relationships. This happens, in short, because agents, and people, often \emph{want} to filter by race, whether they are conscious of that fact or not \cite{anderson2014political}. Thus, removing this filter makes it harder to find partners that agents desire. The attribute intervention causes this decrease for a separate, non-obvious reason. Specifically, including additional attributes makes the search space becomes more diverse, thus making it harder to find the ``right'' partner. Most obviously, with more attributes, the chance of two agents matching on all attributes decreases.

Finally, with respect to the norm intervention, influencing the preferences of agents by changing on-platform norms produces fluctuating evaluation criteria, especially when paired with racially biased offline norms. Agents enter the simulation with their own potentially racially-biased preferences. As they interact on the platform, their preferences slowly converge towards the racially unbiased on-platform norm. However, when going through the offline relationship phase, their preferences again adapt to the racially biased societal (offline) norms, thus shifting the evaluation criteria of agents again. These fluctuations means it takes longer to find suitable partners; so, if the process extends over the tolerance of agents for either bad recommendations or failed relationships, they will exit the platform without entering a long-term relationship. In our model, as in modern cultural sociological theory \cite{patterson_making_2014}, then, the impact of the norm intervention is therefore contingent on the extent to which norms within the context of the platform permeate to other normative context by way of individual preferences.  


\subsection{Effects on User Satisfaction}

Our simulations suggest that most interventions, on average, will decrease user satisfaction, especially filter interventions that do not allow users to filter on race. However, the magnitude, and in select cases, even the sign of these effects vary according to the settings of the other interventions, and across the two metrics considered. These findings present evidence that interventions can potentially be carried out without decreasing user satisfaction, but that these interventions typically are weaker (in terms of reducing racial homogamy), that effects on one satisfaction metric may contradict results on another, and that which combinations of interventions are most effective is not obvious a priori.

\begin{figure}[t]
\begin{subfigure}{.95\linewidth}
  \centering
  \includegraphics[width=\linewidth]{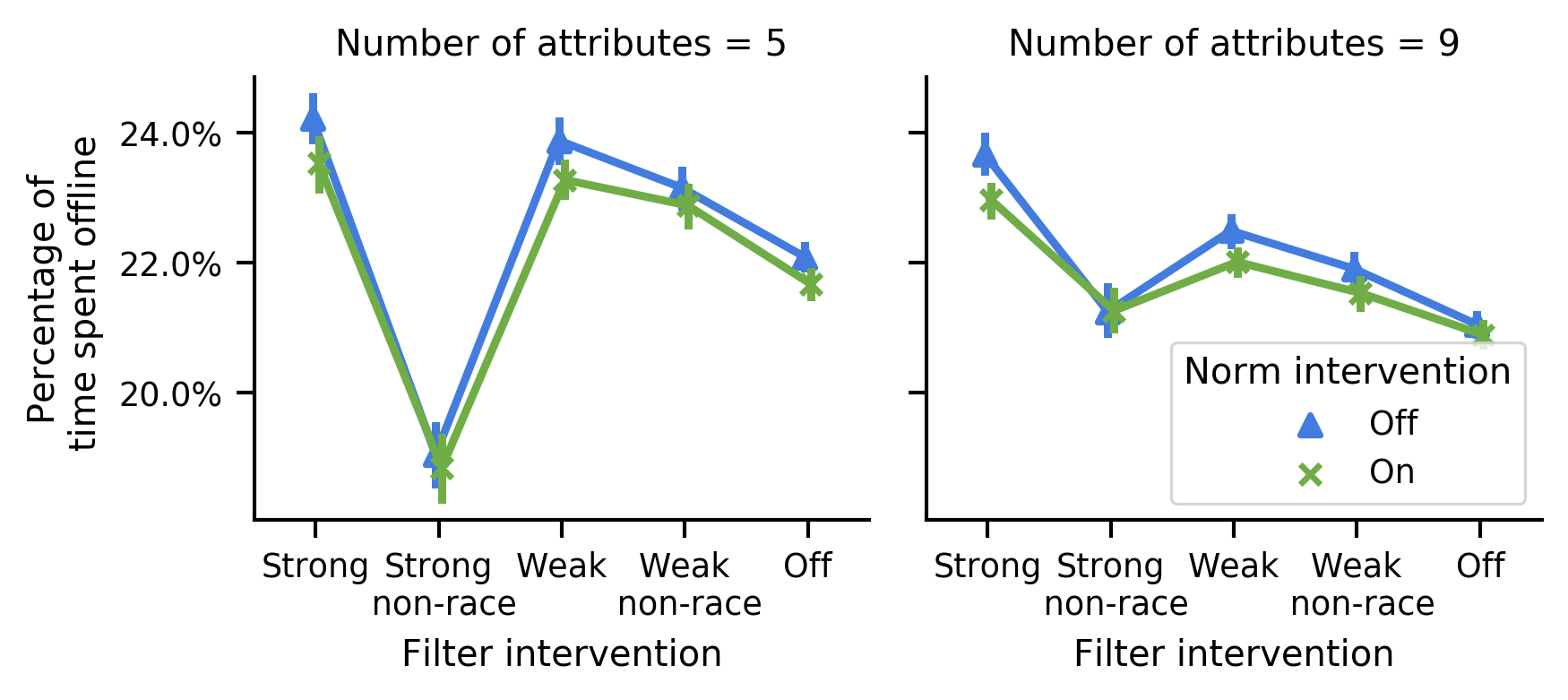}  
  \caption{The percentage of time spent in offline relationships.}
  \label{fig:time_offline}
\end{subfigure}
\begin{subfigure}{.95\linewidth}
  \centering
  \includegraphics[width=\linewidth]{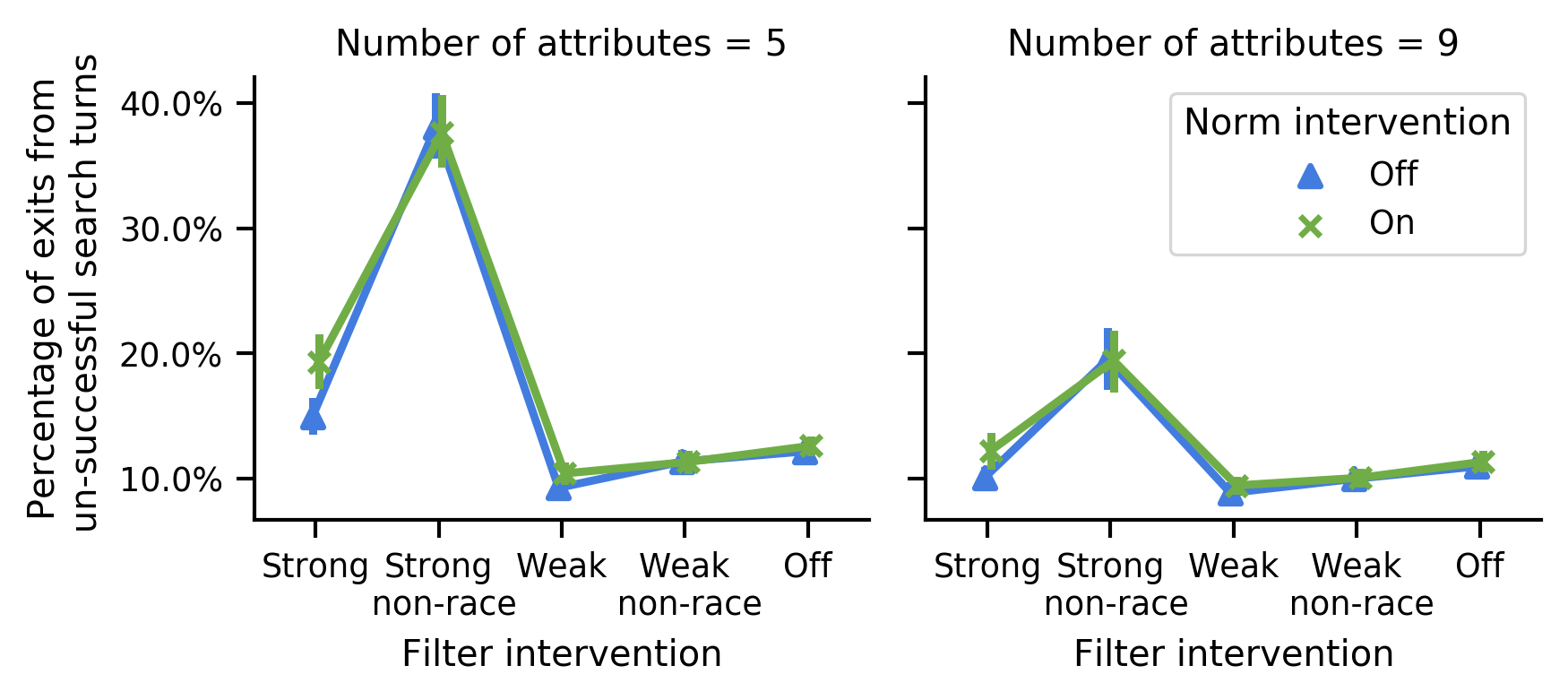}  
  \caption{The percentage of exits due to exhausting the tolerance for un-successful search turns.}
  \label{fig:bad_rec}
\end{subfigure}
\caption{Plots showing the level of user satisfaction with the application of different interventions.}
\label{fig:user_sat}
\end{figure}

Specifically, Figure ~\ref{fig:time_offline} shows that filtering interventions, to varying degrees, all decrease the percentage of agent's time spent in offline relationships, and in doing so likely decrease user satisfaction. The most drastic impacts are from the Strong non-race intervention, where 19.0\% of agents' time is spent offline, as opposed to 24.2\% in the baseline condition.  This effect is, however, mitigated when paired with the attribute intervention. This is likely because agents become more satisfied with the non-race filtering options they have available to them.   Other interventions, while less capable of reducing racial homogamy than the Strong non-race filter, have a much more limited effect on user satisfaction. 

These observations for time spent offline also appear when measuring user satisfaction in term of the percentage of users who leave the platform without a partner. As shown in Figure~\ref{fig:user_sat}, however, certain filter interventions can actually \emph{increase} user satisfaction along this metric. Specifically, the use of Weak or no filtering can increase satisfaction on this metric, because agents often do not report all of their attributes, and thus including also the profiles with unknown attribute-values lowers the chance of excluding possibly interesting partners.  Full descriptions of other patterns in Figures~\ref{fig:time_offline} and \ref{fig:user_sat} are included in the Supplemental Material.


%

\subsection{Intervention Impacts under Different Assumptions about Society}

\begin{figure}[h]
    \centering
    \includegraphics[width=\linewidth]{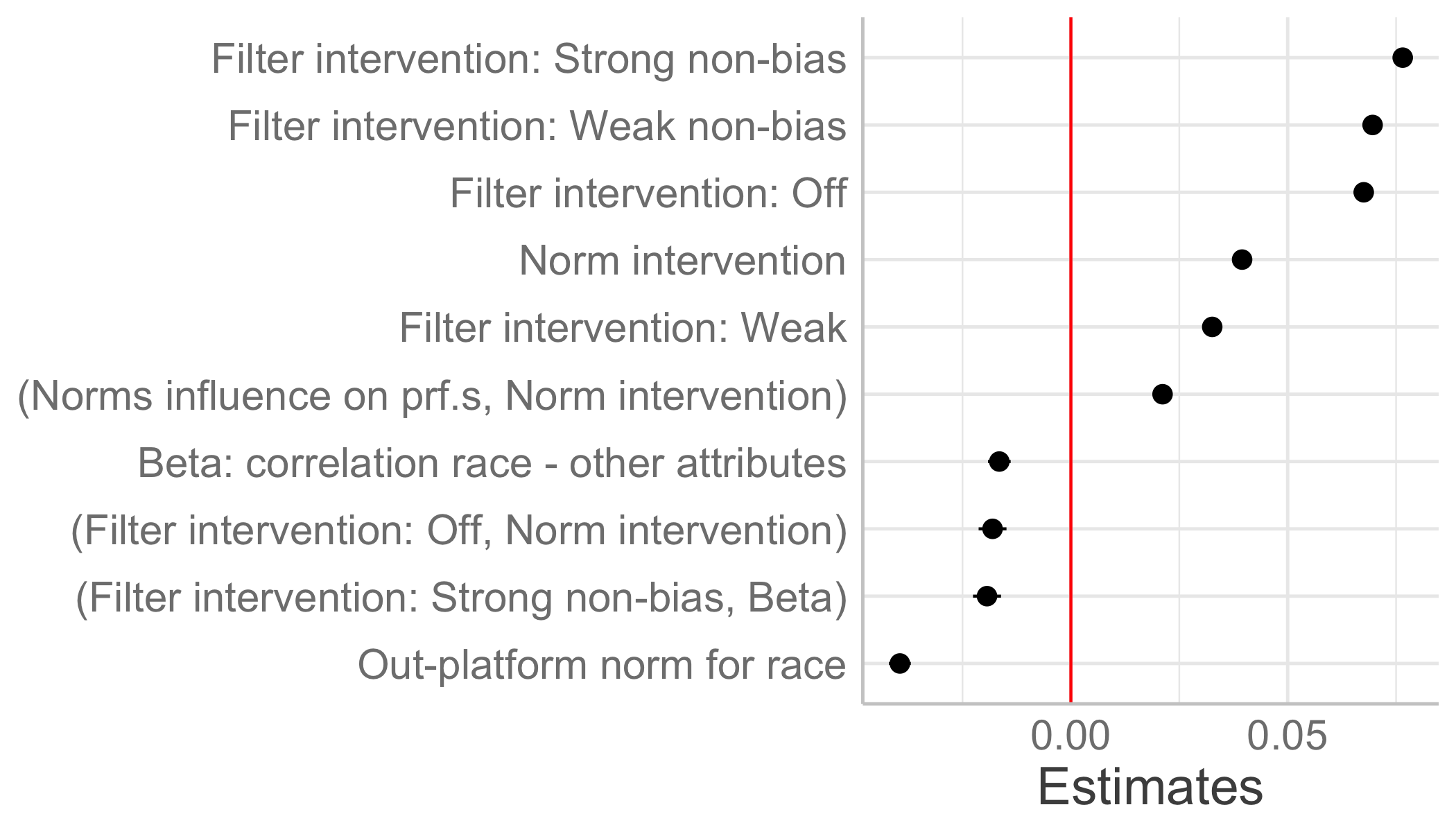}
    \caption{The first and second order marginal effects of the varied parameters on the percentage of out-group long-term relationships, for the 10 effects of strongest absolute value. }
    \label{fig:2nd_dep}
\end{figure}

Finally, we explore how interventions fare under different assumptions about society.  Broadly speaking, we find, as reflected in Figure~\ref{fig:2nd_dep}, that interventions do have strong stand-alone beneficial effects, but that their impact is diluted if certain, often reasonable, assumptions about society are met. 

The figure shows coefficients resulting from a linear regression with racial heterogamy as the outcome variable, and all (centered and scaled) model parameters in Table~\ref{tab:virt_exp} \emph{and their pairwise interactions} as independent variables. For readability, Figure~\ref{fig:2nd_dep} only includes the 10 coefficients with the strongest absolute effect. The regression model fits the data quite well ($R^2 = 0.83$), indicating that first- and second-order effects explain much of the dynamics of the complex system we model, but also that the model cannot be fully explained by phenomenon enumerable by theory alone.\footnote{See the Supplementary Material for full results of the first-order effects.}

Filter interventions have the largest absolute effect on racial heterogamy. On average, across all simulated societies, removing the possibility to filter by race increases racial heterogamy by $7.65\%$ (CI = $(7.43, 7.88)$)\footnote{Note that confidence intervals are provided here but should be cautiously interpreted, as they can be made infinitely small simply by running additional simulations}. The Norm intervention also has a considerable positive impact, as it alone induces an average increase of $3.95\%$ (CI = $(3.73,  4.17)$). The impact of these interventions reflects their potential to have real and possibly lasting changes on society.

Not all interventions we tested, however, increase racial homogamy in the manner theorized by Hutson et al. \cite{hutson2018debiasing}. Specifically, we find that only \emph{matching} (as opposed to competing) and \emph{searchable} (as opposed to experiential) attributes produce a meaningful increase in racial heterogamy. Adding other types of attributes can actually have negative effects, either in decreasing overall numbers of relationships while not impacting the number of interracial relationships, or in simply leading to a decrease in racial heterogamy.\footnote{See the Supplement for full results} These findings underscore the importance of using simulation to think through the operationalization of potential interventions, especially by combining various theoretical models (here, of intervention and attribute types) into a single simulation model.

Other results in Figure~\ref{fig:2nd_dep} show, unsurprisingly, that different assumptions about society reflect drastically different baselines from which this improvement might come. For example, we see that racial heterogamy is significantly \emph{reduced} as correlations between race and the other attributes increase, and with an increase in racial bias in off-platform norms.  Perhaps most importantly, however, Figure~\ref{fig:2nd_dep} reveals that intervention effects and societal conditions can \emph{interact} to change the effectiveness of an intervention. For example, the Norm intervention is particularly effective when we make the assumption that agent preferences are strongly impacted by social norms. But, adding a non-race filtering option is less effective with increased values of the $\beta$ parameter. Put another way, the effectiveness of not allowing individuals to search by race on an online platform is limited by the extent to which race is signaled by, or correlated with, other attributes that the platform \emph{does} allow one to search on.

\section{Conclusion}

In their recent articles, \citet{hutson2018debiasing} and \citet{anderson2014political} present contrasting views on the extent to which online dating platforms can intervene to impact long-term patterns of racial homogamy in online dating. In the present work, we developed an agent-based modeling approach that synthesizes modern perspectives in (cultural) sociology, the psychology of online dating, and social computing to help us explore this debate. 

Before discussing the implications of our work, we again explicitly call out two critical assumptions that our model makes. While we allow race to co-vary with other individual attributes, reflecting its multidimensional nature \cite{sen_race_2016}, our virtual experiment does ultimately assume that race, and gender, are both binary, i.e. that one can easily and completely specify an individual's race or gender. Doing so lends an aura of racial and gender determinism to our work that FAccT scholars are rightfully skeptical of  \cite{hanna2020towards}. Our experiment also assumes that these variables are independent, and thus ignores important intersectional identities.  However, we emphasize that these assumptions are made only in the present virtual experiment, and are by no means hard-coded into our open source simulation framework. We hope that we and others can therefore easily explore other constructions of race, dating, gender, and sexual preference in future work.

These limitations in mind, our findings shed critical light on the effectiveness and side effects of the interventions proposed by \citet{hutson2018debiasing} in reducing racial homogamy, and ultimately racial inequality. Our work implies that the extent to which these interventions will be effective depends on four things.  First, it depends on what we assume about society. We find that some of the proposed assumptions are effective only if we assume that people are unable to infer (perceived) race from other cultural preferences along which race can be performed online, an assumption that is potentially untenable given what is known about correlations between cultural attributes in general \cite{dellaposta_why_2015} and perceptions of race in particular \cite{freeman_dynamic_2011}. And, other interventions are effective only if we assume that 1) the platform can significantly impact social norms towards interracial dating on their site and that 2) these norms have more significant effects on individuals than other ``offline'' normative structures (e.g. the family) that have historically encouraged racial homogamy. 

Second, it depends on what platforms, and society, are willing to accept as sacrifice for progress on racial homogamy. Our model makes explicit these potentially non-obvious tradeoffs, quantifying how certain interventions to reduce racial homogamy have negative impacts on user satisfaction and rates of overall long-term relationships.  Third, it depends on how long one is willing to wait. We find that changing on platform norms have a limited effect; however, our model only studies short-term effects. Changes in large-scale social norms can take much longer, and thus the payoff of these normative interventions may exist only much further into the future as on-platform norms influence agent preferences, which in turn slowly reshape norms in other contexts. Finally, it depends on what one's ultimate goal is. If one's ultimate goal is to increase racial homogamy to any degree, our results provide evidence that this is possible under a fairly wide range of assumptions.  However, if one's goal is to eradicate racial homogamy, then interventions at the platform level seem to hit a saturation point well below true equality. These technological interventions will therefore almost certainly never be enough; they must be supplemented with structural and normative social change beyond the platform. 

While one could argue that these findings are post-hoc somewhat obvious, our work is motivated by a disagreement in the existing literature over the potential for platform intervention. This should serve as an important reminder that everything is obvious once it is observed \cite{watts2011everything}, and we hope motivates the use of our approach to other domains. Especially when unsuitable changes could have large negative effects, constructing a model and simulating different possible outcomes could help explore potential problems to those interventions that emerge within a complex sociotechnical system.

\begin{acks}
We thank Yuhao Du, David Garcia, Atri Rudra, Christoph Stadtfeld, Aleksandra Urman, and the FAccT reviewers for their valuable feedback on this work. KJ was supported by an NSF award IIS-1939579, with partial support from Amazon.
\end{acks}

\bibliographystyle{ACM-Reference-Format}
\bibliography{refs}


\end{document}


\title{An Agent-based Model to Evaluate Interventions on Online Dating Platforms to Decrease Racial Homogamy}
\title[\resizebox{4.2in}{!}{An Agent-based Model to Evaluate Interventions on Online Dating Platforms to Decrease Racial Homogamy}]{An Agent-based Model to Evaluate Interventions on Online Dating Platforms to Decrease Racial Homogamy}
\subtitle{Supplemental Material}

\author{Stefania Ionescu}
\email{ionescu@ifi.uzh.ch}
\orcid{}
\affiliation{%
  \institution{University of Zürich}
  \streetaddress{Andreasstr. 15}
  \city{Zürich}
  \state{Switzerland}
  \postcode{8050}
}

\author{Anikó Hannák}
\email{hannak@ifi.uzh.ch}
\orcid{}
\affiliation{%
  \institution{University of Zürich}
  \streetaddress{Andreasstr. 15}
  \city{Zürich}
  \state{Switzerland}
  \postcode{8050}
}

\author{Kenneth Joseph}
\email{kjoseph@buffalo.edu}
\orcid{}
\affiliation{%
  \institution{University at Buffalo}
  \streetaddress{Davis Hall}
  \city{Buffalo}
  \state{NY, USA}
  \postcode{14260}
}

\maketitle

\renewcommand{\shortauthors}{Trovato and Tobin, et al.}

\appendix
\section{Further Details on the Decision Model}

Recall that the decision function is based on the following equation: 
$$\mathbb{P}(\text{yes}) = \frac{1}{1 + \exp \left(- \mathbb{E} \left[w^{(i)}\cdot s^{(i)}(a) |  \hat{a}^{(j)} \subseteq a \right]- t \right)}.$$

 Formally, $\mathbb{E} \left[w^{(i)}\cdot s^{(i)}(a) |  \hat{a}^{(j)} \subseteq a \right] = \sum_{a \subseteq \hat{a}^{(j)}} \left( w^{(i)}\cdot f(a^{(i)}, a) \right) \cdot \mathbb{P}_{G^{(i)}} (A = a'|\hat{a}^{(j)}\subseteq A)$, where $\mathbb{P}_{G^{(i)}} (A = a'|\hat{a}^{(j)}\subseteq A)$ to be the probability an agent with the known attributes $\hat{a}^{(j)}$ has the vector of attributes $a'$ in the view of agent $i$. This probability is obtained by dividing the entry for $a'$ in $G^{(i)}$ over the sum of the entries corresponding to all possible extension of $\hat{a}^{(j)}$.
 
\subsection{Attribute Scoring Function}

In the main text, we informally introduced the attribute scoring function. Below, we include its formal definition:
\[ 
\left(s^{(i)}(a)\right)_k= \left\{
\begin{array}{ll}
      \ 1 & \text{,if } k \text{ matching and } a^{(i)}_k = a_k \text{, or } \\
       & k \text{ competing and } a^{(i)}_k < a_k \\
      \ 0 & \text{,if } k \text{ competing and } a^{(i)}_k = a_k\\
      -1 & \text{,if } k \text{ matching and } a^{(i)}_k \neq a_k \text{, or } \\
       & k \text{ competing and } a^{(i)}_k > a_k \\
\end{array} 
\right. 
\]

\subsection{Time Variable}
Finally, note that the time contribution gets reset to $0$ when moving from an online to an offline relationship. This is consistent with the literature, as the first offline date is a crucial point in deciding whether or not the relationship will continue \cite{finkel2012online}.

\section{Further Details on Initialization}

\subsection{Agent attributes}

To generate agent attributes during initialization of a new agent, we use a generative model common in both the decision-making literature \cite{cantwell2020thresholding} and to other areas of computational social science, e.g. topic modeling \cite{blei2007correlated}. For each agent $i$, we start by generating one instance, $d^{(i)}$, of a $K+1$ dimensional normally distributed random variable $A \sim \mathcal{N}(\mu, \Sigma)$. For interpretability, we use a zero-mean distribution, i.e. we fix $\mu = 0$. Therefore, $d^{(i)}$ will be the degree to which agent $i$ fits with a $1$ category on each attribute. For example, if $d^{(i)}_0 = 0.5$ then $i$ identifies more as having a race-value of $1$, while if $d^{(i)}_0 = -1$ they identify more as having a race-value of $0$. Moreover, without loss of generality, we fix the variance of each component to $1$, i.e. we let $\Sigma_{j, j} = 1$. To complete the values of the covariance matrix, $\Sigma$, we use two parameters: 1) a parameter $\beta$ capturing the level of correlation between the protected attribute and the others, and 2) a parameter $\gamma$ capturing the level of correlation between the non-protected attributes. That is, $\Sigma_{0, j} = \beta$ and $\Sigma_{j, t} = \gamma$ for every $j \neq t$ non-zero. Varying these two parameters changes the predictability of one attribute, given another. In our virtual experiment, both $\beta$ and $\gamma$ are changed to reflect various degrees of correlation (none, low, medium, and high) between the protected attributes and all other attributes ($\beta$), and between all non-protected attributes (low and medium) for $gamma$ - see Table~\ref{tab:virt_exp}).
The binary attribute vector for each agent is obtained from the signs of the entries in $d^{(i)}$: negative values of $d^{(i)}_j$ correspond to $a^{(i)}_j = 0$, while non-negative value correspond to $a^{(i)}_j = 1$. 
Also generated from this distribution is the \emph{contingency table}, $G$, which is used by agents to construct their \emph{stereotypes}.

\subsection{Stereotypes}
Agents do not necessarily know the true contingency table $G$. Instead, based on their previous experience, they form a belief over this table, namely $G^{(i)}$ for agent $i$. We refer to this as the personal stereotypes of agents \cite{krueger1996personal}. This matrix is generated together with a new agent, and, as mentioned in the main text, changes thought their time in the simulation. In this subsection we disucss its initialisation.

Since this matrix is a result of the experience of the agent prior to entering the platform, we first create a sample of agents that they encountered (either directly or indirectly) before entering the platform. Because racial homophily\cite{mcpherson_birds_2001} and in-group favoritism\cite{hastorf_they_1954,tajfel_social_1971} influence the formation of stereotypes, this sampling is not race-balanced. We first generate, using the true contingency table $G$, $300$ agents having the same race as agent $i$ and $100$ having a different race. Next, similar to previous work \cite{joseph_coevolution_2014}, we use the degree of ethnocentrism, $e$, and alter the true attributes of the different-race agents in the second sample. More precisely, for each agent of a different race in the sample, and each attribute that is perceived as good by agent $i$ there is a chance $e$ of changing it into a negatively-perceived attribute.

The (modified) sample of $400$ agents induces a belief for agent $i$ over which attribute combinations are more likely to be encountered. Formally, this belief is captured into the stereotype matrix $G^{(i)}$.

\subsection{Norms and preferences}

As mentioned in the main text, out-platform norms, agent preferences, and on-platform norms are initialized sequentially. 

To start, the out-platform norm vector is generated inside the $K$th simplex. 
More precisely, since there is a restriction on the value on the race-component and the sum of searchable attributes versus the sum of experiential attributes, three parts are generated separately. 
The first part corresponds to the race attribute and is a $(K+1)$-dimensional vector with the first value given by the respective parameter and 0s on all the remaining positions. We refer to this as $n^{off}_{race}$.
The second part corresponds to the searchable attributes. The sum on the components corresponding to the searchable attributes is given by accounting for the weight of the race attribute and for the relative importance of searchable vs experiential attributes, i.e. this part is given by the equation $\text{sum}_{\text{searchable}} = 1 - (\text{the norm value on the race attribute})\times (\text{the importance of searchable attributes})$. Next we take an $S$- dimensional vector uniformly at random from the $(S-1)$-th simplex, complete it with $(K+1-S)$ values of $0$ (one on the first position, and the reaming at the end), and multiply it by $\text{sum}_{\text{searchable}}$. This gives the searchable part of the out-platform norm, namely $n^{off}_{searchable}$.
The last part, $n^{off}_{experiential}$, is obtained similarly as the searchable one. The out-platform norm is the sum of these three components, i.e. $n^{off} = n^{off}_{race} + n^{off}_{searchable} + n^{off}_{experiential}$.

Next, the individual preferences are obtained by taking the out-platform norm and adding some noise.
 Last, the on-platform norm vector is obtained by averaging the preferences on the searchable components.
  More precisely, for each agent $i$ on the platform we consider $w^{(i, s)}$ to be the searchable entries of the preference vector of that agent. Since $w^{(i)}\in \Delta^{K}$, the sum of the searchable entries of that vector might be below $1$, say $s_i = \sum_j w^{(i, s)}_j$. A re-scaling of $w^{(i, s)}$ will ensure that it is in the $(S-1)$th simplex, therefore $w^{(i, s)}$ becomes $\frac{1}{s_i} \cdot w^{(i, s)}$. We refer to this version of $w^{(i, s)}$ as the searchable vector of preferences. Their average is the initial on-platform norm, i.e. $n^{on} = \frac{1}{N} \cdot \sum_{i\leq N} w^{(i, s)}$. 
  
  When intervening on the on-platform norm, its value on the protected characteristic, i.e. its first entry, remains 0. To reflect this, after initially setting the on-platform norm as the average of preferences, we let its first entry to be $0$, and re-scale the others to keep the norm in the $(S-1)$th simplex. Similarly, after getting the norm closer to the average of preferences, we once agin set its first entry to $0$ and re-scale the other ones. With such an intervention in place, at the end of each iteration, when preferences are getting closer to norms, the agents' weight on the protected characteristic will lower.

 \section{Further Details on Updates}
 \subsection*{Norms and preferences}
 We now turn our attention to norm and preference updates, which, as mentioned in the main text, are carried out in sequence during each iteration. First, the preferences update after each interaction to reflect its outcome. Second, at the end of each iteration, the on-platform norm gets closer to the average preference of agents. Third, the preferences are updated by getting closer to norms. Each of this three updates are done by taking convex combinations between the original value of the norms/preferences and the values of the influencing vector. Since both preferences and norms are in a simplex, which is closed under convex combinations, they will remain in the simplex after the update. The remaining of this section explains the details of these updates.

First, the preferences of each agent in a relationship update based on the interaction. Using the decision function, agents decide whether or not to continue the relationship. We consider a mutual continuation decision as a sign of a \textit{positive interaction}, and any other decision combination as a sign of a \textit{negative interaction}. Let agent $A$ update their preferences after an interaction of quality $q$ with agent $B$. In this case, we construct an interaction vector $i$. The preference of $A$ will then update to the convex combination between this interaction vector and the old preference vector.

We construct the interaction vector into stages. First, based on the quality of the interaction and how good each attribute is perceived, we determine whether the weight of each attribute should increase, decrease, or stagnate (see Table~\ref{table:change_interaction}). For example, if there was a negative interaction with an agent having a good attribute, then perhaps that attribute is not so important, and its weight should decrease. Second, for each attribute $k$ where the value should decrease, we put $i_k = 0$, and for each attribute $k$ where the value should stagnate we put the previous weight value, i.e. $i_k = w^{(A)}_k$. Last, the interaction values for attributes that whose weight should increase are completed in proportion to their old-value, such that $i\in\Delta^K$. As a note, we only construct such a vector if there is some relative change, i.e. some attribute weight values should increase and some should decrease. If, for example, there is some positive interaction with somebody with all attributes perceived as being bad, then there will be no post-interaction preference update, as this indicates no relative change in preferences. Using this interaction vector, the preferences of agent $A$ updates to a convex combination between $i$ and the old preference, with a strength given by some parameter $\theta$, i.e. $\theta \cdot i +(1-\theta) \cdot w^{(A)}$.

\begin{table}[h!]
\centering
\begin{tabular}{c | c c c} 

 & $k$ good & $k$ neutral/unkwnon & $k$ bad \\ 
 \hline\hline
 $q$ = positive & $\uparrow$ & $-$ &$\downarrow$ \\ 
 $q$ = negative & $\downarrow$ & $-$  & $\uparrow$ \\

\end{tabular}
\caption{Table showing whether the interaction vector should dictate an increase, decrease, or stagnation in each attribute $k$ depending on the value of the interaction $q$.}
\label{table:change_interaction}
\end{table}

Next, at the end of each turn, the on-platform norms are changed to get closer to the average of the searchable preferences of agents. As a reminder from Section~\ref{sec:gen}, the searchable vector of preferences is obtained by taking from the weight vector of preferences only the searchable entries, and re-scaling it to be in the $(S-1)$th simplex. The on-platfom norms are updated to the convex combination between the average of agents' searchable vector of preferences and the old value of on-platform norms. Lastly, afterwords, each preference becomes the convex combination between its old value and the norm. Depending on whether the agent is online or offline, the normsin question could be the on-platform norm, which updates only the searchable attributes, or the out-platform norm.

\section{Further Details on Fixed Parameter Values}
As outlined in the table showcasing the parameters in our model, some parameters remain constant thought the experiment. In some cases, their values are chosen based on past work, while for others we did a sensitivity test to see if they have a meaningful change. Below, we discuss in turn some of the reasoning that went behind fixing each of these parameters:
\begin{itemize}
    \item \textit{Strength of updates: interaction $\to$ preferences.}
    
    In the initial experiment we also tried more extreme values, such as a $50\%$ and an $100\%$ influence. Increasing this parameter also increases the unpredictability of the results: the variance and the time taken to achieve convergent behaviour surge. Therefore, we settled for more reasonable values on this parameter.
    \item \textit{Strength of updates: preferences $\to$ norms.}
    
    For this parameter, we also tested larger values, such as $10\%$, $20\%$ and $50\%$. Using such values did not affect the results, especially when paired with low level of influences of interactions on preferences, i.e. when preferences were more stable.
    
    \item \textit{Relative importance of out-platform norms when generating initial preferences.}
    This parameter controls how much alike the preferences of agents are originally, and how much they resemble the out-platform norms. Variations of these parameter influenced more the short-term results, i.e. the ones after only a couple of hundred of iterations, rather then the end-results observed after $2000$ iterations
    
    \item \textit{Number of initial acquaintances in-group and out-group.}
    
    We have also tried having less such acquaintances, namely $150$ and $50$. Besides a slight increase in variability no difference was observed.
    \item \textit{Weight of searchable versus experiential attributes.}
    
    \citet{frost2008people} conducted a study on how much weight do searchable attributes have when compared to experiential attributes when considering a potential partner. We choose the values of this parameter based on their result.
    \item \textit{The probability a searchable attribute is specified on the profile and the probability of learning an attribute of someone else.}
    
    Past work indicates that users do not always specify all the attributes they could on their dating profile\cite{lewis2016preferences} and that attributes are discovered during interactions \cite{finkel2012online}. These motivates the using of these two parameters in our model. The exact numbers influence only the amount of time the two agents act on incomplete information. Small variations to these numbers do not qualitatively change our results.
    
    \item \textit{The number of rounds until a relationship becomes offline and long-term and the frequency of offline interactions.}
    The time it takes one to move to the next step of their relationship as well as how often they interact varies a lot across people. Therefore, we used values close to the averages reported in previous work to set these parameters \cite{ramirez2015online, brym2001love, munoz2007aggression}.
    
    \item \textit{The relative time spent looking at received messages versus new profiles during search and number of profiles and messages considered per iteration.}
    During their time searching, users can both look at profiles and consider messages \cite{finkel2012online}. Based on previous work, we estimated that, on average, one considers around $30$ potential partners (either for sending an initial message, or for giving a reply to such a message and start an online relationship) \cite{frost2008people, finkel2012online}.
    
    \item \textit{Time-span for observing the platform.}
    The simulation runs for $2000$ iterations, the rough equivalent of $5$ years. In our sensitivity test we also tracked $5000$ iterations, but the results revealed that $2000$ are enough to fully observe the effects of interventions, i.e. achieve convergence and small variance.
    
    \item \textit{Initial population size and the number of agents added per iterations.}
    The sensitivity test also experimented with using less ($150$ initially and $2$ added per iteration). This decrease did not produce qualitative changes. Due to computational limitation we did not further increase the number of agents.
    
    \item \textit{Types of agents.}
    In the main text, agents can be of two types, loosely mimicking a heteronormative online dating website. However, we also tested a platform with only one type of agent, thus moving form a two-sided matching problem to a one-sided matching one. The results remind qualitatively similar.
\end{itemize}

Further analysing the impact of changing these parameter could potentially reveal other interesting phenomena. For example, we know that users usually meet within $1-2$ weeks, but what would happen if an intervention was made such that they are encouraged to meet sooner? That being said, this is outside the scope of our paper and is thus left for future research. Even though such an analysis could reveal new dependencies between parameters as well as potentially new effective interventions, they should not change the causal implications observed in this paper.

\section{Additional Results and Result Details}

\subsection{Explaining Patterns in User Satisfaction Metrics}

Norm interventions produce smaller and often insignificant drops in user satisfaction, as they change the underlying preference of users, so also the way they perceive potential partners. The slight drop, of only 1.06\% on average, in offline-time when intervening on on-platform norms is due to the persistence of off-platform norms: even though users change their preference while on the platform because of the intervention there, they will still go through an offline state, when their preferences are altered to reflect these different norms. When increasing the number of attributes, the negative effects of adding non-race filtering options decrease. For example, a shift to Strong non-race changes the percentage of time spent offline from 23.6\% to 21.2\%. The reason for this is that removing the possibility to search by one attribute affects users less if they have multiple others to search by; with more attributes there is a higher chance of having a second attribute that is almost as important as the first.

Slightly different results can be noted with respect to the dissatisfaction resulting from un-successful search turns (see Figure 5b in the main text). While the addition of non-race filtering options on top of both strong and weak filtering increase the level of un-satisfaction, the adding of variety with weak filtering and sometimes even with no filtering does not necessarily do so. Since agents often do not report all of their attributes, including also the profiles with unknown attribute-values lowers the chance of excluding possibly interesting partners. Consequently, when around 50\% of users do not report their value on the search attribute, a strong filtering option excludes many potentially suitable candidates, thus explaining the gaps. Similarly to our other measures of satisfaction, disregarding of the filtering or norm intervention, the addition of attributes improves the satisfaction. When agents have more attributes to search on, excluding one of these or indirectly influencing the preferences through norms disturbs less the capabilities of users to signal and search based on this desires. 

\subsection{Regression Model with First-Order Effects}

\begin{figure}[h]
    \centering
    \includegraphics[width=.9\linewidth]{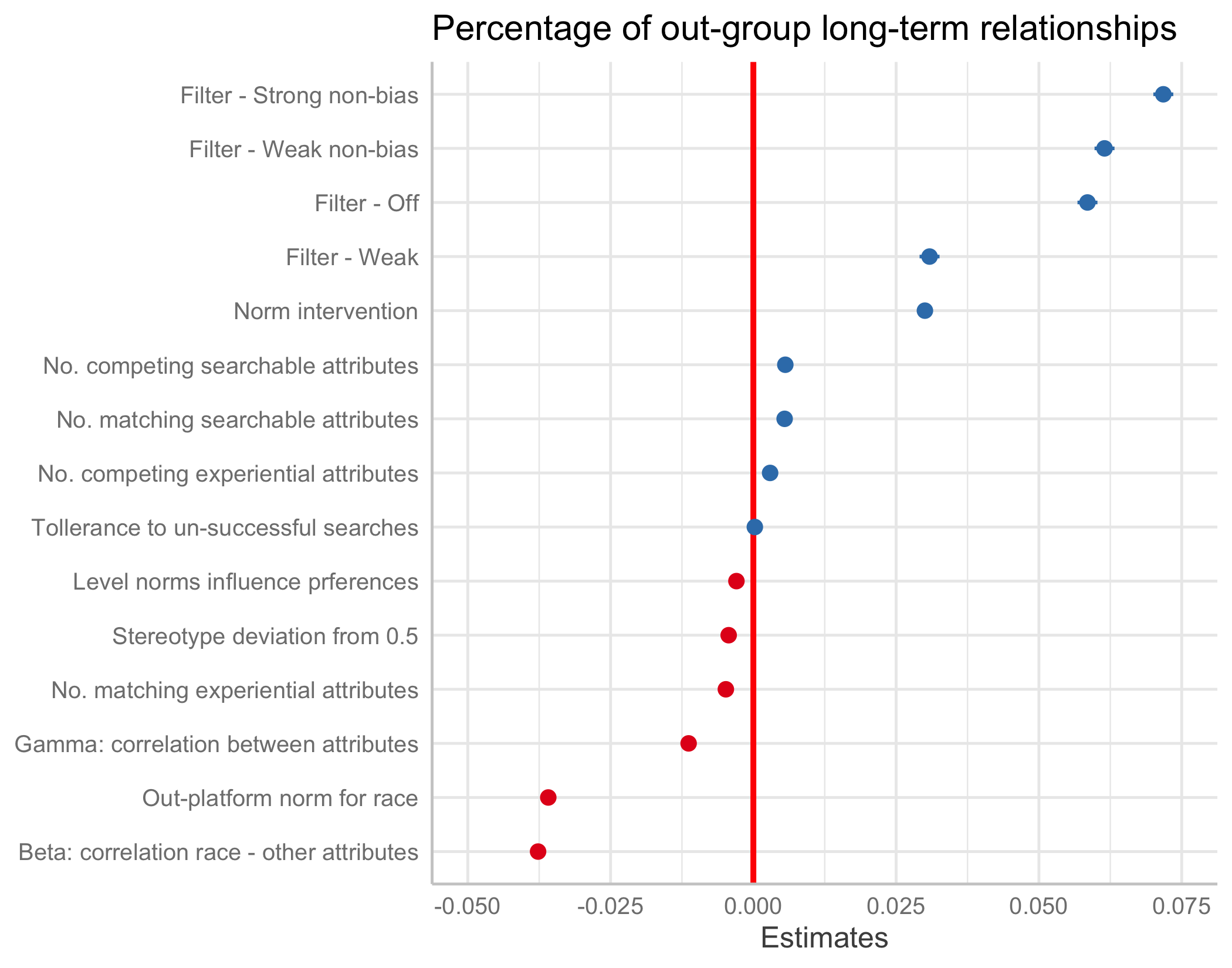}
    \caption{The marginal effect of each varied parameters on the percentage of out-group long-term relationships. }
    \label{fig:1st_dep}
\end{figure}

In the main text, we used a linear regression with racial heterogamy as the outcome variable and the varied model parameters and their pairwise interactions as independent variables to analyse the impact of parameters and pair of parameters on the percentage of out-group long-term relationships. That model had a high predictability ($R^2 = 0.83$). However, the first-order terms alone also fit data quite well ($R^2 = 0.79$). In this part of the supplementary material we look at all the first order effects that were not already discussed in the main text. Figure \ref{fig:1st_dep} provides an overview of these effects.

Before, we have seen that racial heterogamy is significantly \emph{reduced} as correlations between race and the other attributes increases, i.e. $\beta$, and with an increase in racial bias in off-platform norms. Figure \ref{fig:1st_dep} shows that the correlation between non-race attributes also negatively impacts heterogamy. 


In contrast, as expected, there are also societal structures that can \emph{increase} racial heterogamy. For example, adding more attributes usually improves the percentage of out-group relationships, with the only exception being the addition of matching experiential attributes. To understand why this is the case, one needs to remember that the effect on the percentage is a mix between the effect on the total number of long-term relationships and the effect on the number of out-group ones. Adding competing attributes decreases the total number of relationships and has no effect on the number of out-group relationships. Consequently, adding them increase the percentage. Conversely, adding matching attributes has no effect on the total number of relationships, so the effect on the percentage is given by the change in the number of out-group ones. Increasing the number of matching attributes that are observable while online result in more out-group relationships, while adding matching attributes that are only observable when offline produce a decrease. In conclusion, even though adding attributes generally increases the percentage of out-group relationships, the only attributes that produce a meaningful improvement are the matching searchable ones, as their effect is due to the increase in the total number of out-group long-term relationships  rather than due to a decrease in the overall number of relationships.

The level of stereotyping is another interesting parameter, as values away from $50\%$, in either direction have a negative effect on the percentage of out-group relationships. 
Putting this another way, stereotyping that agents of a different race have half of their "good" attributes "bad" have worse effect than both perceiving the different race agent as they are and believing that all of their attributes are "bad" (where "good" and "bad" matching attributes are in relationship with the agent considering them).
To give an example, the blue agent in Figure~\ref{fig:rel_phases}, has a higher chance of finding a partner within a different bias group if either they believe that people with a race of $1$ have exactly the true distribution over music liking ($0\%$ stereotyping) or that they never like music ($100\%$ stereotyping) than if they think half of the ones that do actually like music and have a different race do not have a passion for it ($50\%$ stereotyping). To understand why this is the case, let us consider the case when the blue agent observes another agent with a music passion and an un-specified race. If the blue agent has a $100\%$ level of stereotyping they will assume that the agent has the same race and possibly also associate other positive attributes with them. When they learn the true race value of the agent they will also have learned more of the true attributes of the agent, so by that time, race might not be a determining factor in decision making. 
On the other hand, if the blue agent has a $50\%$ level of stereotyping, the blue agent still perceives them being of a different race as a possiblity, which might result in projecting the perceived negative attributes on the agent in question, consequently resulting in a lower probability of forming and continuing a relationship.

Two smaller effects are due to the level at which norms influence preferences and the tolerance level of agents to un-successful search turns. Although their marginal effect on the percentage of out-group relationships is small, they affect the user satisfaction. For example, an increased level of influence of the norms on preferences results in higher dissatisfaction when implementing norm interventions due to an increase in the drop rate after the first relationships, while a lowered bad recommendation tolerance decreases both the time spent offline and the percentage of exits due to too many un-successful search turns.

\subsection{Minimal Group Paradigm Intervention}
\begin{figure}[h]
    \centering
    \includegraphics[width=\linewidth]{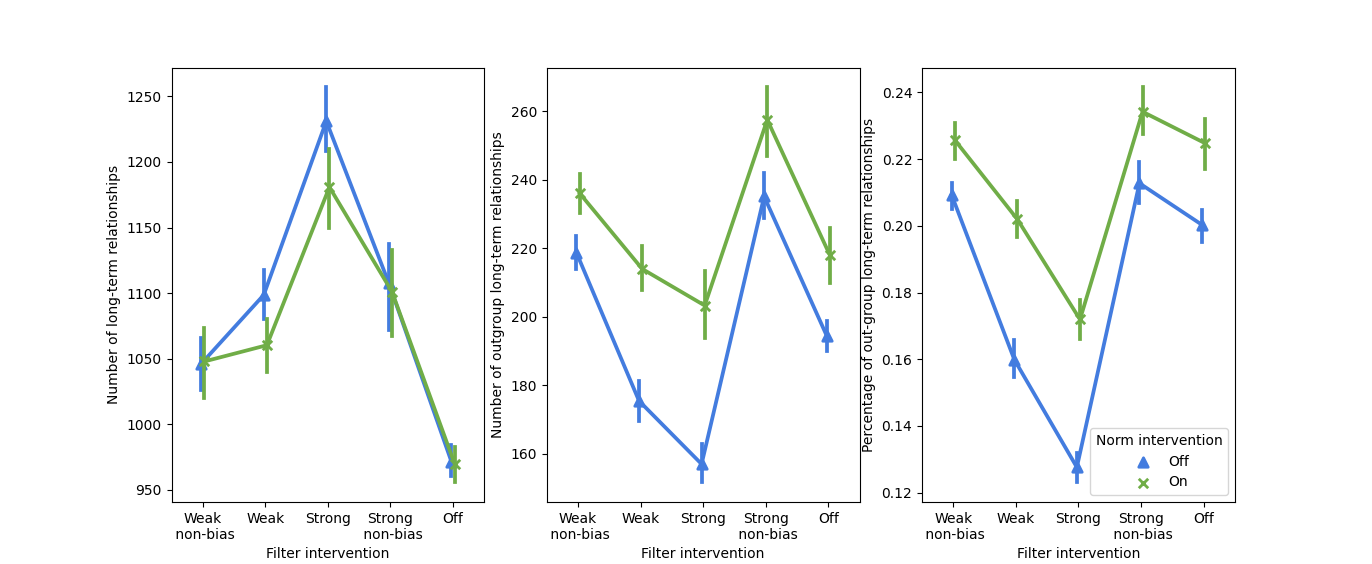}
    \caption{The percentage and number of out-group long-term relationships after the minimal group paradigm intervention.}
    \label{fig:minimal_gr_par}
\end{figure}

As a variant of the attribute intervention, \citet{hutson2018debiasing} also suggest creating attributes that help users "categorize potential partners along new axes unassociated with protected characteristics". This intervention is based on on the minimal group paradigm in social psychology \cite{tajfel_social_1971}. To implement it, we added a new attribute searchable and matching attribute that is correlated neither with race nor with any other preexisting attributes. We also made the assumption that agents value this newly introduced attribute in the same way they value the others. That is, when generating and updating preferences and norms we treat these attribute as any of the other matching searchable non-bias attribute. Figure \ref{fig:minimal_gr_par} shows the impact of this intervention on racial heterogamy as well as on the total and out-group number of long-term relationships.

\begin{figure}[h]
    \centering
    \includegraphics[width=\linewidth]{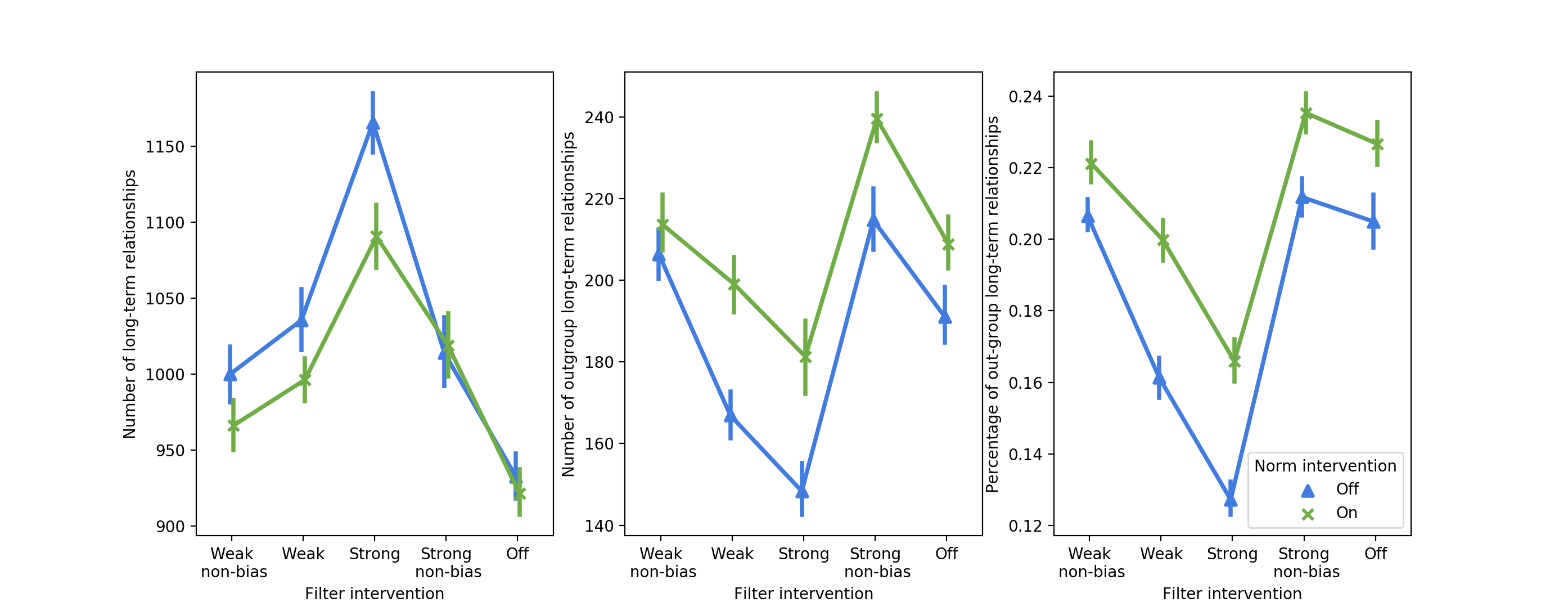}
    \caption{The percentage and number of out-group long-term relationships after the attribute intervention.}
    \label{fig:attr_intvn_2}
\end{figure}

Similarly to the attribute number intervention, this intervention produces an increase in the percentage of out-group long-term relationships over the baseline (see main text for this figure). Moreover, under the assumption mentioned before, i.e. that creating this new attribute can be done in such a way that users value it as much as the other attributes, this intervention is more effective that the general attribute intervention (Figure \ref{fig:attr_intvn_2}). More precisely, the increase in the percentage of out-group long-term relationships is the same, but the decrease in the total number of relationships is smaller and the increase in the number of out-group ones is larger. The analysis of the first order effects in the section above explains this difference. As observed there, matching searchable attributes are the only type of attributes with meaningfully effects, i.e. an improvement in heterogamy due largely to an increase in the number of out-group long-term relationship rather than to a decrease in the total number of relationships. Thus, one woud expect the minimal group paradigm intervention, which only adds matching searchable attributes, to have more positive effects than the general one, which adds one attribute of each.

\bibliographystyle{ACM-Reference-Format}
\bibliography{refs}